\pdfoutput=1
\documentclass[10pt]{IEEEtran}  
\usepackage{amsmath,amssymb,amsfonts}
\usepackage{graphicx}
\usepackage{cite}
\usepackage{algorithmic}
\usepackage{textcomp}
\usepackage[table,dvipsnames]{xcolor}
\usepackage{hyperref}
\usepackage{booktabs}
\usepackage{mathtools}
\usepackage{multirow}
\usepackage{subfig}  
\usepackage{booktabs}
\usepackage{enumitem}
\usepackage{subcaption}
\usepackage{comment}
\graphicspath{{PNG/}} 
\usepackage{etoolbox}
\AtBeginEnvironment{table}{\footnotesize}
\usepackage[normalem]{ulem} 

\def\BibTeX{{\rm B\kern-.05em{\sc i\kern-.025em b}\kern-.08emT\kern-.1667em\lower.7ex\hbox{E}\kern-.125emX}}

\usepackage[none]{hyphenat} 
\usepackage{microtype}      
\begin{document}

\title{Learning Interior Point Method Central Path Projection for Optimal Power Flow}

\author{Farshad Amani, \textit{Graduate Student Member, IEEE}, Amin Kargarian, \textit{Senior Member, IEEE}, Ramachandran Vaidyanathan, \textit{Senior Member, IEEE}
\vspace{-15pt}
\thanks{This work was supported by the National Science Foundation under Grants ECCS-1944752 and ECCS-2312086.

The authors are with the Department of Electrical and Computer Engineering, Louisiana State University, Baton Rouge, LA 70803 USA (e-mail: famani1@lsu.edu, kargarian@lsu.edu, vaidy@lsu.edu).
}
}

\maketitle
\begin{abstract}
This paper proposes a learning-based approach to accelerate the interior-point method (IPM) for solving optimal power flow (OPF) problems by learning the structure of the IPM central path from its early stable iterations. \uline{Unlike traditional learning models that attempt to predict the OPF solution directly, our approach learns the structure of the IPM trajectory itself, since even accurate predictions may not reliably reduce IPM iterations.} The IPM follows a central path that iteratively progresses toward the optimal solution. While this trajectory encodes critical information about the optimization landscape, the later iterations become increasingly expensive due to ill-conditioned linear systems. Our analysis of the IPM central path reveals that its initial segments contain the most informative features for guiding the trajectory toward optimality. Leveraging this insight, we model the central path as a time series and use a Long Short-Term Memory (LSTM) network to project the path using only the first few stable iterations. To ensure that the learned trajectory remains within the feasible region—especially near the optimal point—we introduce a grid-informed mechanism into the LSTM that enforces key operational constraints on generation, voltage magnitudes, and line flows. This framework, referred to as Learning-IPM (L-IPM), significantly reduces both the number of IPM iterations and overall solution time. To improve generalization, we use a sampling-based strategy to generate a diverse set of load conditions that effectively span the operational space. Simulation results across a range of test systems—including a 2869-bus European transmission network—demonstrate that L-IPM achieves up to a 94\% reduction in solution time and an 85.5\% reduction in iterations, without compromising feasibility or accuracy. We also compare L-IPM with classical solution-prediction-based warm starts and show that our central-path projection strategy offers superior performance.

\end{abstract}


\section*{Nomenclature}
\addcontentsline{toc}{section}{Nomenclature}

\begin{description}[leftmargin=2.8cm,labelwidth=2.4cm,labelsep=0.3em,font=\normalfont]

\item[\emph{Indices and Sets:}]
\item[$i,j$] Indices for buses
\item[$d$] Indices for loads
\item[$t$] Iteration step
\item[$\mathbb{B}$] Set of all buses
\item[$\mathbb{G}$] Set of generator buses
\item[$\mathbb{L}$] Set of transmission lines
\item[$\mathbb{D}$] Set of loads

\item[\emph{System Parameters:}]
\item[$Y_{ij}$] Entry of the bus admittance matrix
\item[$F_{ij}^{\max}$] Maximum loading of line $(i,j)$
\item[$P_{G_i}^{\min}, P_{G_i}^{\max}$] Generator power limits at bus $i$
\item[$V_i^{\min}, V_i^{\max}$] Voltage magnitude limits at bus $i$
\item[$a_i, b_i, c_i$] Cost coefficients for generator at bus $i$

\item[\emph{Optimization Variables:}]
\item[$p_{i}, q_{i}$] Active/reactive power generation at bus $i$
\item[$v_i$] Complex voltage at bus $i$
\item[$\theta_i$] Voltage angle at bus $i$
\item[$s_i$] Complex power injection at bus $i$
\item[$f_{ij}$] Apparent power flow on line $(i,j)$

\item[\emph{Slack, Barrier, and Dual Variables:}]
\item[$s_i^{+}, s_i^{-}$] Slack variables for generator limits
\item[$v_i^{+}, v_i^{-}$] Slack variables for voltage limits
\item[$\sigma_{ij}$] Slack variable for transmission line limits
\item[$\mu$] Log-barrier parameter in IPM
\item[$\lambda_i^S$] Multiplier for power balance at bus $i$
\item[$\eta_i^{+}, \eta_i^{-}$] Multipliers for generation limits
\item[$\zeta_i^{+}, \zeta_i^{-}$] Multipliers for voltage constraints
\item[$\xi_{ij}$] Multiplier for thermal line limit on $(i,j)$

\item[\emph{GI-LSTM Variables:}]
\item[$x_t$] System state at iteration $t$
\item[$y_t$] Predicted OPF solution at time $t$
\item[$C_t$] LSTM cell state
\item[$h_t$] LSTM hidden state
\item[$f_t, i_t, o_t$] Forget, input, and output gates in LSTM
\item[$k_{\text{max}}$] Number of IPM iterations used in training
\item[$\theta$] Trainable parameters of the LSTM
\item[$y_t^{\text{LSTM}}$] Predicted solution from GI-LSTM
\item[$y_t^{\text{final}}$] Final OPF solution after refinement

\item[\emph{Loss Function Terms:}]
\item[$L_{ACOPF}$] grid-informed loss function for ACOPF
\item[$L_{DCOPF}$] grid-informed loss function for DCOPF
\item[$\omega_1, \omega_2, \omega_3$] Weights balancing prediction accuracy and constraint violation

\item[\emph{Data Generation and Sampling:}]
\item[$P_{d}$] Active load at bus $i \in \mathbb{D}$
\item[$P_{d}^{\text{nominal}}$] Nominal load at bus $i \in \mathbb{D}$
\item[$\alpha_d$] Load scaling factor
\item[$n_d$] Number of load buses
\item[$n_{\text{total}}$] Total number of load combinations
\item[$n_{\text{samples}}$] Number of Latin Hypercube samples
\item[$X_i^{\text{OPF}}$] OPF solution for sample $i$
\item[$\bar{X}^{\text{OPF}}$] Mean OPF solution over samples
\item[$\sigma_X^2$] Mean squared deviation in OPF results
\item[$Z$] Normal distribution critical value
\item[$\epsilon$] Tolerated estimation error
\item[$\sigma$] Standard deviation 

\end{description}
\section{Introduction}
\subsection{Background}
\IEEEPARstart{E}{ffective} management of economic dispatch, voltage regulation, and system security is key to reliable grid operation, which makes optimal power flow (OPF) an essential tool in modern power systems \cite{mohammadian2025restoring, pagnier2022machine}. As power systems become more complex, the need for fast and scalable OPF solutions becomes more pressing. With the increasing reliance on renewable energy sources, combined with the unpredictable nature of loads and distributed generation, real-time OPF becomes even more challenging to compute. To keep up with these demands, it is necessary to have OPF solvers that are computationally efficient and practical for real-world power systems \cite{siano2012real}. These solvers should surpass existing methods in both speed and scalability.

Several methods have been proposed to accelerate OPF computation and improve its efficiency, ranging from advanced mathematical formulations to emerging paradigms such as quantum computing \cite{mohammadi2016agent, amani2026quantum}. Among recent algorithmic advancements, techniques such as semidefinite programming \cite{wang2018sdp}, quantum-enhanced optimization \cite{amani2024quantum}, sequential quadratic programming \cite{soofi2020socp}, and methods based on the augmented Lagrangian frameworks \cite{jha2022distribution} have shown considerable promise. In addition to these optimization-based approaches, some researchers have explored hardware and parallelization enhancements, such as GPU acceleration and high-performance computing \cite{alawneh2023review}, which can improve performance but require access to specialized infrastructure. Others have used heuristic and metaheuristic algorithms \cite{buch2019efficiency}, which may offer faster solutions but do not guarantee global optimality.

More recently, attention has turned toward machine learning-based methods, including grid-informed neural networks, supervised learning approximations, reinforcement learning, and graph neural networks \cite{khaloie2025review, amani2026event}. Many of these approaches typically reduce problem dimensionality or directly predict the OPF solutions. As a result, they generally provide less direct leverage over the optimizer’s iterative behavior. 
\vspace{-8pt} 
\subsection{Motivation and Research Gap}

Among all OPF solution methods, the interior point method (IPM) remains one of the most popular and reliable techniques \cite{jabr2002primal}. IPM operates by iteratively applying Newton's steps to optimize system variables, where each step involves solving a large linear system of equations. Typically, the most computationally expensive part of the algorithm is solving these linear systems \cite{gondzio2002reoptimization}. Some researchers have explored quantum linear solvers to address this computational burden, which offer potential asymptotic speedups \cite{amani2025optimal}. However, hardware limits, errors, and scalability challenges still prevent their use in large-scale OPF \cite{kaseb2024hybrid}. Other approaches to enhance IPM include heuristic methods such as inexact Newton strategies \cite{venzke2020inexact} and machine learning-based techniques \cite{khaloie2025review}. 

Existing machine learning-assisted OPF efforts can be broadly grouped into three categories: (i) direct end-to-end prediction of the solution variables, which may produce erroneous outputs with limited control over feasibility and solution quality; (ii) warm-start approaches that predict high-quality solution guesses to start closer to the optimum, yet for iterative solvers such as IPM, this does not reliably reduce iterations or runtime; and (iii) iteration-reduction or feasibility-oriented schemes that steer iterates toward feasible regions, often improving feasibility but potentially sacrificing optimality~\cite{rozycki2025energy, dai2025temporal, siano2012real}. Among these, warm-start strategies are especially appealing in practice because they can provide accurate initial primal points; however, even accurate initial primal points do not reliably translate into fewer Newton steps or lower runtime. This is because IPM convergence depends not only on the primal variables, but also on the consistent initialization and subsequent evolution of dual variables, slack variables, and barrier/centrality conditions that maintain feasibility and guide the iterates along the central path~\cite{venzke2020inexact}. \uline{In practice, even when IPM starts from a nearly optimal primal solution, it may still need about the same number of iterations, or even more, to restore centrality and satisfy the full Karush–Kuhn–Tucker (KKT) conditions. Case studies show that an optimal OPF warm start does not always reduce the iteration count; in some cases, IPM converges in fewer iterations without it.} In contrast, our objective is to reduce runtime by directly reducing the number of IPM iterations while preserving feasibility. Rather than focusing solely on the initial point, we exploit the iterative behavior of the IPM central path to learn informative update directions and accelerate convergence.

IPM central path follows a trajectory toward the optimal solution, which can be interpreted as a time series, as it progresses iteratively toward convergence. In this work, we propose modeling the trajectory of the central path as a data sequence and learning its behavior using a Long Short-Term Memory (LSTM) network, owing to its ability to capture temporal dependencies in sequential data \cite{khan2023short}. A key characteristic motivating this approach is that the most informative patterns in the IPM central path trajectory emerge during the early iterations, while the final iterations primarily serve to guarantee feasibility and enforce constraint satisfaction. By leveraging this property, we can apply path sampling techniques to project the IPM central path, enabling a significant reduction in the number of Newton steps required and, consequently, a substantial decrease in the overall OPF computation time.
\vspace{-8pt}
\subsection{Contributions}

Applying LSTM to project the central path of IPM presents several challenges. Since OPF is a constrained optimization problem, direct projections without attention to the grid physics and optimization constraints do not inherently guarantee a high-quality solution. Infeasible solutions can lead to constraint violations, compromising system reliability and potentially having catastrophic consequences in real-world power grid operations. To address this, we propose a grid-informed LSTM (GI-LSTM) that incorporates system constraints into the learning process. Another key challenge lies in generating a sufficiently diverse and representative training dataset. For a given power network, the number of possible operating conditions grows exponentially with the number of buses and load variations. Ensuring the generalization and reliability of the learning IPM (L-IPM) approach, therefore, requires careful attention to both data generation and model architecture.

The contributions of this paper are summarized as follows.
\begin{itemize}
    \item The IPM central-path trajectory for OPF is analyzed to extract informative patterns from the early iterations, which capture valuable relationships between grid physics and optimization variables. Building on these insights, we propose an L-IPM framework that leverages early-iteration behavior to project the central path toward a high-quality OPF solution.
   \item To address OPF computational and modeling challenges, the proposed approach reduces the required number of IPM iterations by relying on the early steps, where the underlying matrices are best conditioned and most reliable to compute. In contrast, later iterations tend to become increasingly ill-conditioned, which can degrade numerical reliability and increase computational burden. 
    \item To ensure the solution quality, we incorporate a grid-informed loss function that penalizes violations of generator limits, voltage magnitudes, and power flow constraints.
    \item To capture a wide range of system conditions, we use Latin hypercube sampling (LHS) to generate diverse load scenarios without the need for exhaustive enumeration \cite{iordanis2025efficacy}.
\end{itemize}

Simulations on various systems, including a realistic large 2869-bus system, show that the proposed L-IPM significantly reduces the number of IPM iterations, Newton steps, and solution time while maintaining OPF solution accuracy.

\vspace{-9pt}
\section{Preliminaries}
This section outlines the standard approach for solving the OPF problem, focusing on the widely adopted formulation that uses IPM as its core optimization technique.
\vspace{-6pt}
\subsection{OPF Formulation}
OPF minimizes total generation costs subject to power balance and system constraints \cite{zimmerman2016matpower}: 

\begin{equation}
\min_{\mathbf{p}} \sum_{i \in \mathbb{G}} \left(a_i p_{i}^2 + b_i p_{i} + c_i \right)
\end{equation}
OPF is subject to AC power flow equations that enforce active and reactive power balance at each bus:  
\begin{equation}
s_i = p_i + j q_i = v_i \sum_{j \in \mathbb{B}} Y_{ij}^* v_j^*
\end{equation}
The problem is further constrained by generator, voltage, and transmission line thermal limits in (\ref{genlim}), \ref{vollim}, and (\ref{thermlim}).
{\small
\begin{equation}
\label{genlim}
P_{i}^{\min} \leq p_{i} \leq P_{i}^{\max}, \quad \forall i \in \mathbb{G}
\end{equation}
\begin{equation}
\label{vollim}
V_i^{\min} \leq v_i \leq V_i^{\max}, \quad \forall i \in \mathbb{B}
\end{equation}
\begin{equation}
\label{thermlim}
|f_{ij}| \leq F_{ij}^{\max}, \quad \forall (i,j) \in \mathbb{L}
\end{equation}}
    \vspace{-6pt}
\subsection{Classic IPM}
IPM, as the state-of-the-art method to solve OPF, introduces slack variables to convert inequality constraints \ref{genlim} to \ref{thermlim} into equality constraints. It then constructs a Lagrangian function by incorporating all constraints through dual variables. The method adds a logarithmic barrier term to the objective function to ensure non-negativity. IPM solves the resulting system iteratively using the Newton-Raphson method, where each iteration involves solving a system of linear equations to approach the optimal solution. The Lagrangian function $\mathcal{L}$ is formed by incorporating the equality constraints (power balance) using Lagrange multipliers $\lambda_i^P$ and $\lambda_i^Q$, and the inequality constraints (generator and voltage limits, thermal limits) using slack variables and their corresponding dual variables. The Lagrangian is given by:

{\small
\begin{align}
\mathcal{L} 
&= \sum_{i \in \mathbb{G}} \left( a_i p_{i}^2 + b_i p_{i} + c_i \right) \nonumber \\
&+ \sum_{i \in \mathbb{B}} Re \left\{ \lambda_i^{S*} \left( s_{G_i} - s_{D_i} - v_i \sum_{j \in \mathbb{B}} Y_{ij}^* v_j^* \right) \right\} \nonumber \\
&+ \sum_{i \in \mathbb{G}} \eta_i^{+} (p_{i} + s_i^{+} - p_{i}^{\max}) 
+ \sum_{i \in \mathbb{G}} \eta_i^{-} (p_{i} - s_i^{-} - p_{i}^{\min}) \nonumber \\
\end{align}}

{\small
\begin{align*}
&+ \sum_{i \in \mathbb{B}} \zeta_i^{+} (v_i + v_i^{+} - V_i^{\max}) 
+ \sum_{i \in \mathbb{B}} \zeta_i^{-} (v_i - v_i^{-} - V_i^{\min})  \\
&+ \sum_{(i,j) \in \mathbb{B}} \xi_{ij} \left( |f_{ij}|^2 + \sigma_{ij} - (F_{ij}^{\max})^2 \right)
- \mu \sum_{(i,j) \in \mathbb{B}} \log \sigma_{ij}  \\
&- \mu \sum_{i \in \mathbb{G}} \left( \log s_i^{+} + \log s_i^{-} \right)
- \mu \sum_{i \in \mathbb{B}} \left( \log v_i^{+} + \log v_i^{-} \right) \\
\end{align*}}

The term \(\Re\) ensures the Lagrangian remains real-valued. To solve the Lagrangian formulation, we construct the KKT conditions, which are derived with respect to all primal variables (e.g., \(\mathcal{PD} = \left\{ p_{i},\ q_{i},\ v_i,\ \theta_i \right\}\)), slack variables (e.g., \(\mathcal{SD} = \left\{ s_i^\pm,\ v_i^\pm,\ \sigma_{ij} \right\}\)), and dual variables associated with the constraints. Newton's method is then applied to solve this system iteratively. At each iteration, a linear system is formed with a coefficient matrix known as the KKT matrix. 
\section{Proposed Grid-Informed L-IPM}
\begin{figure}[h]
    \centering
        \captionsetup{font={footnotesize}}
    \includegraphics[width=0.85\columnwidth]{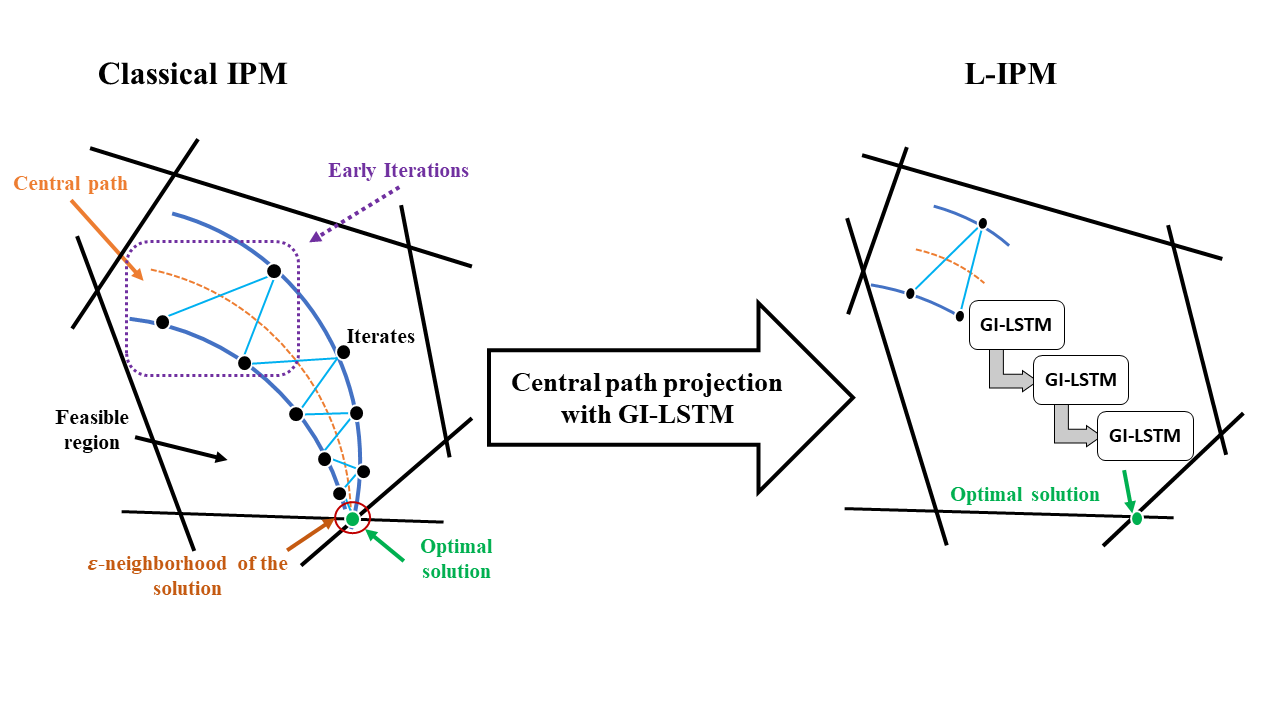}
        \vspace{-20pt}
    \caption{IPM and the proposed L-IPM paths to the optimal solution.}
        \vspace{-10pt}
    \label{IPMpath}
\end{figure}

The IPM-based OPF follows a structured trajectory known as the central path, illustrated in Fig.~\ref{IPMpath}. The initial iterations of the central path contain information about its trajectory and, consequently, the final optimization solution. These early steps typically cover a significant portion of the path toward optimality. The IPM central path closely resembles a time series. Given this characteristic, LSTM networks are well-suited for leveraging early iteration data to predict the solution.

\begin{figure}[h]
    \centering
        \captionsetup{font={footnotesize}}
    \subfloat[Condition number]{\includegraphics[width=0.4\columnwidth]{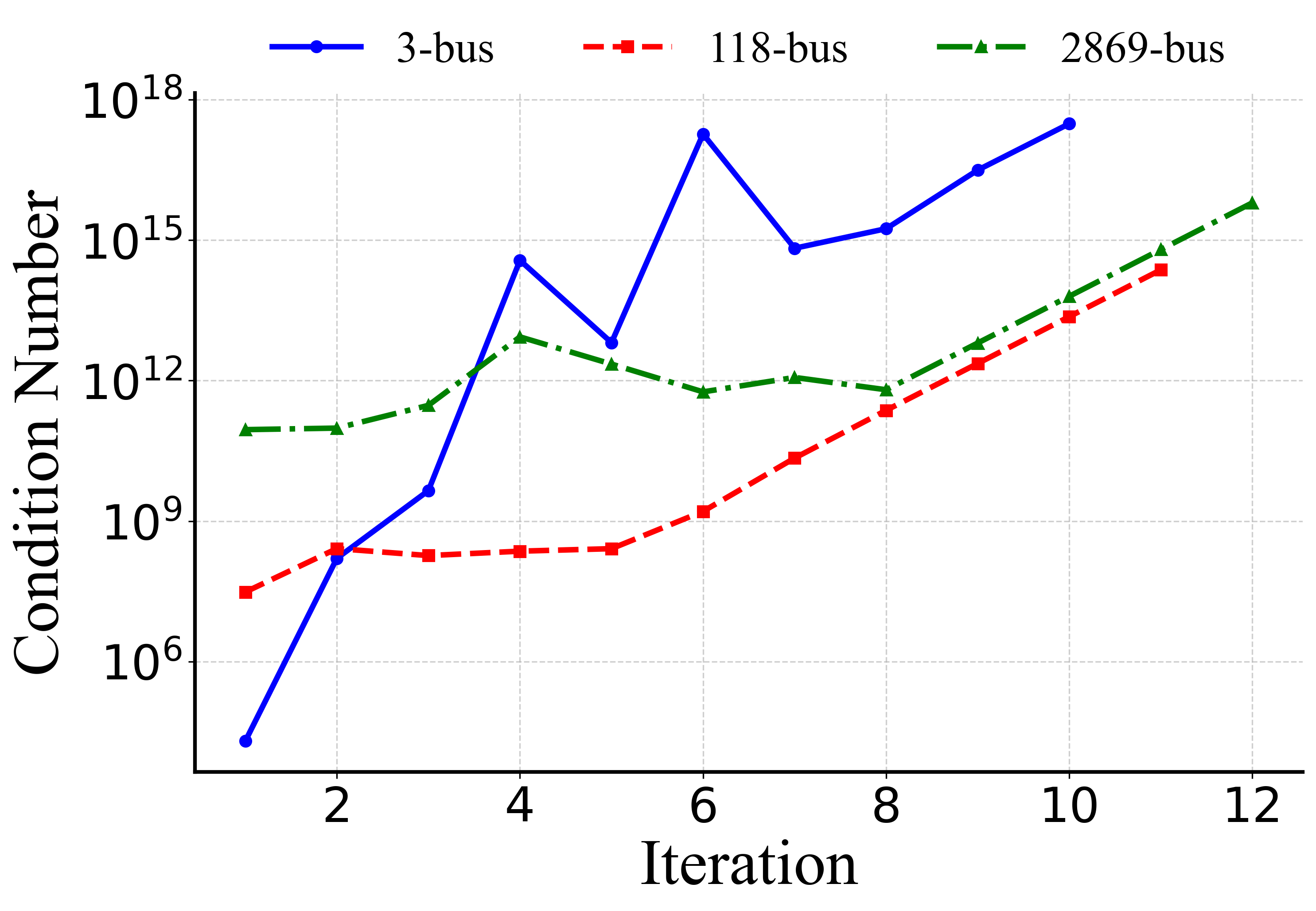}}
    \hfill
    \subfloat[2869-bus PEGAS system]{\includegraphics[width=0.4\columnwidth]{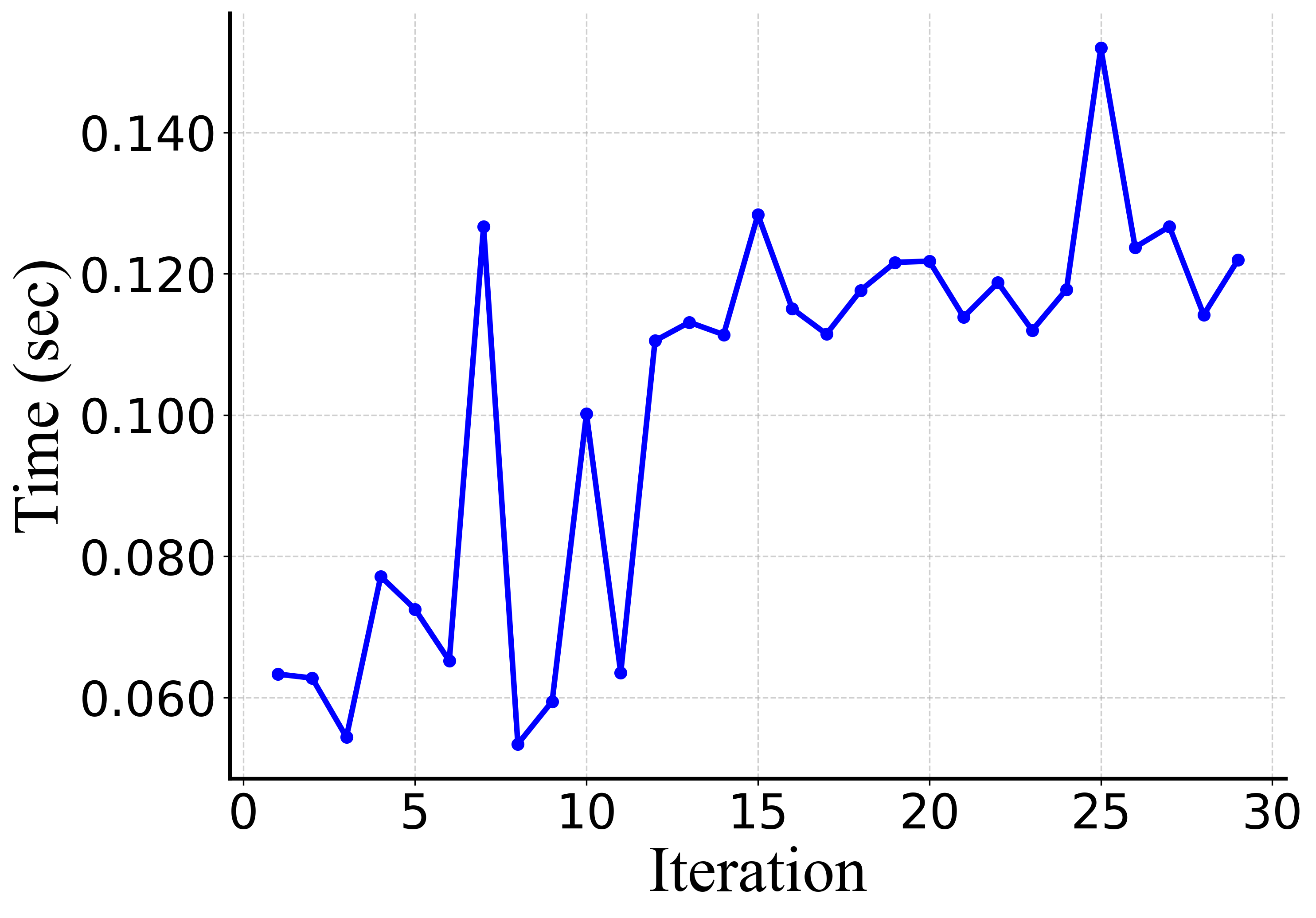}}
    \caption{Condition number and solution time of IPM's iterations in DCOPF.}
        \vspace{-10pt}
    \label{fig:time_cond}
\end{figure}

Furthermore, as Fig.\ref{fig:time_cond}(a), early iterations tend to involve matrices with relatively low condition numbers, making them computationally more stable and efficient to solve. In contrast, the final iterations often involve ill-conditioned matrices, which are more numerically sensitive and generally more expensive to solve. We observed that in many cases, the time needed to solve the IPM iterations increases toward the final iterations, as shown in Fig.~\ref{fig:time_cond}(b). However, a higher condition number does not necessarily lead to longer solution times, as the computational effort also depends on factors such as matrix sparsity, fill-in during factorization, and solver efficiency~\cite{alawneh2023review}. This behavior can also be attributed to the efficient sparse solvers and numerical techniques implemented in MATPOWER. 

\begin{figure}[h]
    \centering
        \captionsetup{font={footnotesize}}
    \includegraphics[width=0.85\columnwidth]{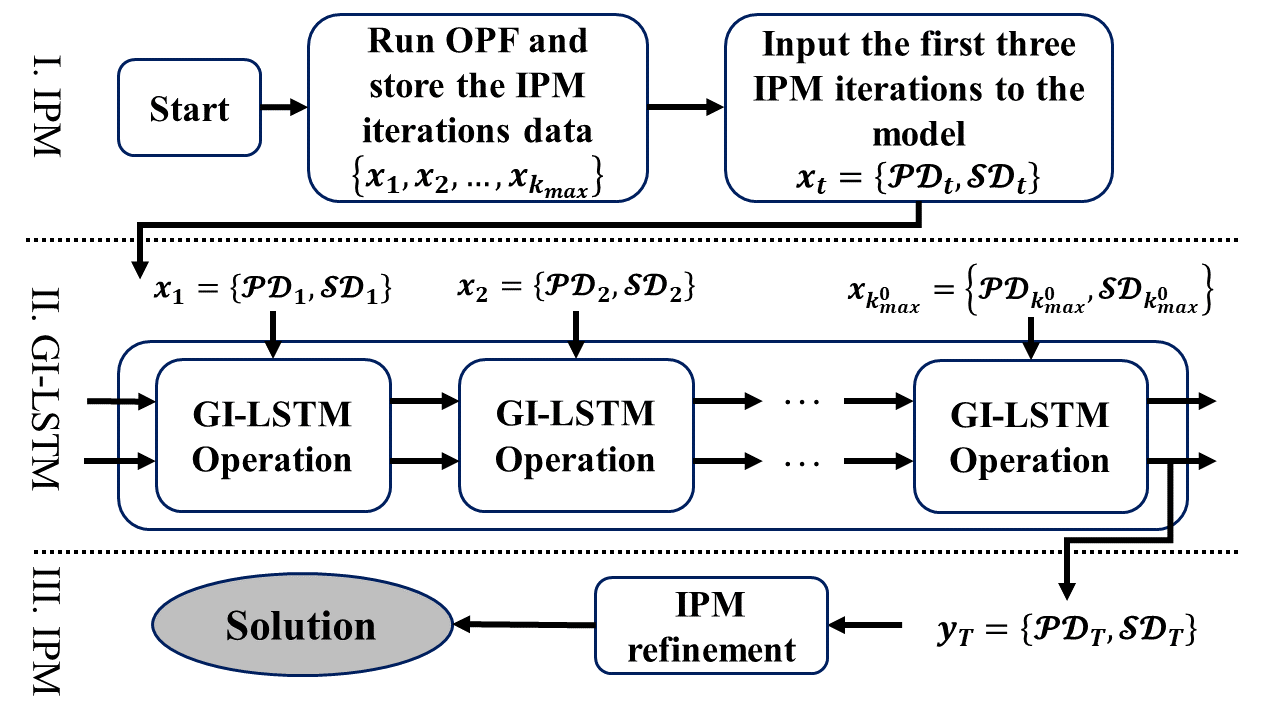}
    \caption{L-IPM framework.}
        \vspace{-10pt}
    \label{LIPM}
\end{figure}

By leveraging early iterations and avoiding direct computation of poorly conditioned steps, we propose an L-IPM approach that significantly reduces the total IPM time required to obtain the OPF solution. The proposed L-IPM framework is illustrated in Fig. \ref{LIPM}.
\vspace{-4pt}
\vspace{-6pt}
\vspace{-6pt}

\vspace{0pt} 
\subsection{Learning Model Architecture}
An LSTM network can be trained to mimic the IPM central path trajectory by learning from the initial iterations obtained through solving the Newton step. Such an LSTM network consists of multiple layers that extract temporal features from past system states, followed by a fully connected layer that maps these features to OPF variables \cite{khan2023short}. 

At each time step $t$, the LSTM cell state updates based on previous iterations:
\vspace{-9pt}
{\small \begin{equation}
C_t = f_t \cdot C_{t-1} + i_t \cdot \tilde{C_t}
\label{7}
\end{equation}
\begin{equation}
h_t = o_t \cdot \tanh(C_t)
\end{equation}}
where $f_t$, $i_t$, and $o_t$ represent the forget, input, and output gates, respectively, controlling the flow of information within the network. The cell state $C_t$ stores long-term dependencies, while the hidden state $h_t$ provides a compressed representation for predicting OPF solutions. For each time step $t$ in (\ref{7}), we solve OPF using \text{MATPOWER} and collect a dataset as structured in Table~\ref{table:dataset_structure}.

\begin{table}[h]
\centering
\caption{GI-LSTM Training Data: Primal--Dual IPM Trajectory (Per Scenario $s$ and iteration $k$)}
\vspace{-5pt}
\label{table:dataset_structure}
\setlength{\tabcolsep}{3.4pt}
\renewcommand{\arraystretch}{1.15}
\begin{tabular}{ccccccc}
\toprule
$s$ & Load & Primal $\mathbf{x}_k$ & Eq. dual $\boldsymbol{\lambda}_k$ & Ineq. dual $\boldsymbol{\mu}_k$ & Feas. \\
\midrule
$1$ & $\mathbf{d}^{(1)}$ & $[P_G,V,\theta,\ldots]$ & $\lambda^{\text{Pbal}},\lambda^{\text{Qbal}},\ldots$ & $\mu^{\text{line}},\mu^{V},\mu^{P_G},\ldots$ & $(\text{Yes})$ \\
$\vdots$ & $\vdots$ & $\vdots$ & $\vdots$ & $\vdots$ & $\vdots$ \\
\bottomrule
\end{tabular}
\end{table}
\vspace{-12pt}
\subsection{LSTM Training Mapping Strategy}
Given a set of system conditions, LSTM is trained to learn a temporal mapping function from a sequence of past states to the final OPF solution \cite{khan2023short}. Formally, the model learns:
\vspace{-5pt}
\begin{equation}
y_t = f_{\theta} (x_{t-k}, x_{t-k+1}, ..., x_t)
\end{equation}

where \( x_t \) represents the system state at time \( t \), including features such as load demand, generation constraints, and historical OPF results, and \( y_t \) denotes the predicted solution. The model parameters \( \theta \) are optimized to minimize the prediction error while encouraging feasible outputs.

In the context of LSTM, this mapping is further specialized to learn the relationship between early-stage iterations of IPM and its final converged solution. Let \( k_{\text{max}}^0 \) denote the number of early IPM iterations used as input, which in our case is 3 iterations. Then, the input-output mapping becomes:
\vspace{-4pt}
\begin{equation}
X = \{ x_1, x_2, \dots, x_{k_{\text{max}}^0} \}, \quad Y = x_T
\end{equation}
\vspace{-2pt}
Each \( x_k \) represents the system state at iteration \( k \), including generator outputs, bus voltages, Lagrange multipliers, and slack variables. The LSTM model learns a function \( f_{\theta}(X) = Y \), where the parameters \( \theta \) are defined as the model's learnable parameters.
\vspace{-8pt}
\subsection{Grid-Informed LSTM for OPF Feasibility}
\begin{figure}[h]
    \centering
        \captionsetup{font={footnotesize}}
    \includegraphics[width=0.4\columnwidth]{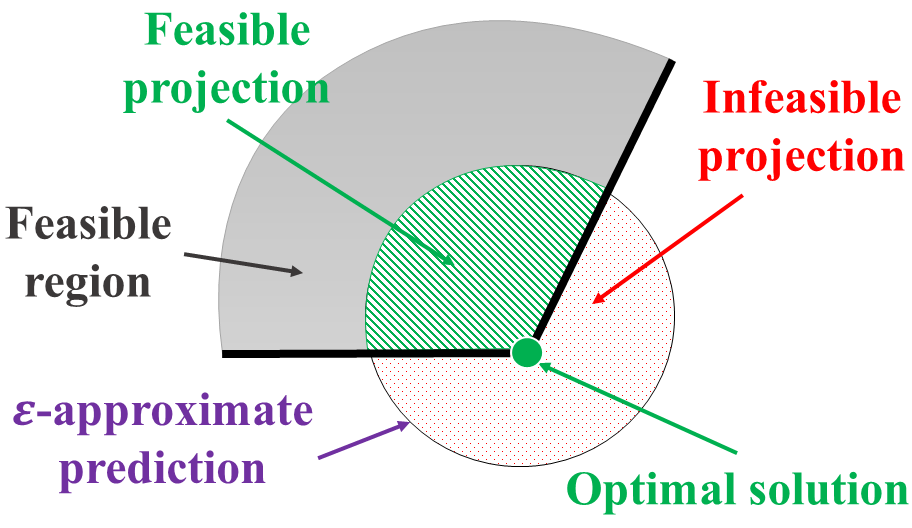}
    \caption{GI-LSTM projection of IPM central path.}
        \vspace{-10pt}
    \label{FeasInfeas}
\end{figure}

The LSTM network projects the final iteration of the IPM or the optimal OPF solution. This is illustrated in Fig. \ref{FeasInfeas}, which provides a zoomed-in view of the optimal solution shown in Fig. \ref{fig:time_cond}(a). While LSTM is effective in learning the convergence trajectory of IPM, ensuring that the LSTM projection of the IPM central path remains feasible or stays within the green region surrounding the optimal solution is critical for maintaining OPF feasibility and satisfying power system operational constraints. If the projection falls within the red zone, it would lie outside the OPF feasible region, thus violating the operational constraints. This is problematic because the proposed L-IPM approach relies on a learning-model framework that uses the learner’s central path projection to find the optimal solution. If the learner’s projection is infeasible and falls outside the feasible region, the last IPM step does not verify the optimality of the solution.

Several key challenges must be addressed to enhance the robustness of L-IPM, including capturing the nonlinear convergence behavior of IPM, maintaining predictive accuracy, and ensuring generalization across diverse operating conditions. We address these challenges using the proposed IPM-tailored GI-LSTM framework.

A standard mean-squared error loss optimizes prediction accuracy but does not ensure physically feasible OPF solutions. To enforce feasibility, a grid-informed loss function for ACOPF and DCOPF is introduced in (\ref{L_ACOPF}) and (\ref{L_DCOPF}), respectively:

{\small
\begin{equation}
\begin{aligned}
L_{ACOPF} = &\ \omega_1 \sum_{i \in \mathbb{G}} \left(P_{G,i}^{\text{LSTM}} - P_{G,i}^{\text{IPM}}\right)^2 + \\
    &\ \omega_2 \sum_{i \in \mathbb{B}} \left(V_i^{\text{LSTM}} - V_i^{\text{IPM}}\right)^2 + \\
    &\ \omega_3 \sum_{(i,j) \in \mathbb{L}} \max\left(0, F_{ij}^{\text{LSTM}} - F_{ij}^{\max}\right)
\end{aligned}
\label{L_ACOPF}
\end{equation}
}

{\small
\begin{equation}
\begin{aligned}
L_{DCOPF} = &\ \omega_1 \sum_{i \in \mathbb{G}} \left(P_{G,i}^{\text{LSTM}} - P_{G,i}^{\text{IPM}}\right)^2 + \\
    &\ \omega_2 \sum_{i \in \mathbb{B}} \left(\theta_i^{\text{LSTM}} - \theta_i^{\text{IPM}}\right)^2 + \\
    &\ \omega_3 \sum_{(i,j) \in \mathbb{L}} \max\left(0, P_{ij}^{\text{LSTM}} - P_{ij}^{\max}\right)
\end{aligned}
\label{L_DCOPF}
\end{equation}
}
Hyperparameters $\omega_1$ to $\omega_3$ balance accuracy and feasibility. The first two terms $\omega_1$ and $\omega_2$ ensure accurate predictions of generator outputs and voltage magnitudes, while the third penalizes power flow constraint violations. Among these, $\omega_3$ is typically assigned the highest value to strongly discourage infeasible solutions.
\vspace{-12pt}
\subsection{GI-LSTM Tuning}
\subsubsection{Number of GI-LSTM Layers}
A two-layer GI-LSTM architecture effectively captures IPM iteration sequences for small- to medium-scale OPF problems and larger networks under typical conditions. For large-scale OPF cases with more complex IPM convergence trajectories, a three-layer GI-LSTM is recommended to better model these dynamics. This configuration balances modeling depth and computational efficiency.
\subsubsection{Regularization Techniques}
To prevent overfitting and enhance generalization across diverse OPF scenarios, we use dropout and L2 regularization. Dropout randomly deactivates neurons during training, with the rate tuned to balance robustness and information retention, while L2 (weight decay) penalizes large weights to simplify the model and improve generalization \cite{khan2023short, amani2025learning}.
\vspace{-12pt}
\subsection{L-IPM for OPF Solution} 
Reducing the number of IPM iterations before predicting the optimal solution allows for substantial computational savings. However, it is important to ensure that the GI-LSTM model generalizes well to diverse OPF conditions. Using L-IPM, which combines GI-LSTM predictions with IPM refinement, we enforce the feasibility and accuracy of the central path projection and, consequently, the OPF solution. The process begins by training GI-LSTM on historical OPF solutions obtained from IPM. Once trained, GI-LSTM generates the central path projection $y_t^{LSTM}$ for the OPF solution, which is then the optimality is verified using a final IPM optimization step:

\vspace{-8pt}
{\small \begin{equation} 
y_t^{final} = \arg\min_{y} || y - y_t^{LSTM} ||^2 \quad \text{s.t. OPF constraints}
\end{equation}}
\vspace{-2pt}
The proposed L-IPM framework modifies the classic IPM algorithm to achieve faster convergence toward the same solution. Feasibility and accuracy are enforced through GI-LSTM and the final IPM verification steps, as illustrated in Fig.~\ref{LIPM}.
\vspace{-12pt}
\section{Data Generation Techniques}

The L-IPM model's effectiveness depends on the training dataset's diversity and representativeness, particularly in capturing the impact of load variations on the power system. The key objective is to determine the minimum number of load scenarios required to ensure generalization while avoiding excessive computational burden. We suggest a systematic approach to estimate the required number of load cases based on power system size, load variability, and OPF sensitivity.
\vspace{-8pt}
\subsection{Defining Load Scenarios and Their Variability}

Let \( \mathbb{D} \subseteq \mathbb{B} \) denote the subset of load buses where active and reactive power demand varies. The power system exhibits different operating conditions due to changes in load levels. These variations are captured using a probabilistic model:
\begin{equation}
    P_{d} = P_{d}^{\text{nominal}} \cdot \alpha_d, \quad \forall d \in \mathbb{D}
\end{equation}
where \( P_{d} \) is the active power demand and \( P_{d}^{\text{nominal}} \) is the nominal demand at bus \(d\). The scaling factor \( \alpha_d \) is modeled as a random variable drawn from a uniform distribution \( U(a,b) \), where the bounds \( (a,b) \) depend on the considered load scenario. Specifically, \( \alpha_i \sim U(0.9,1.1) \) for the nominal load condition, \( \alpha_i \sim U(1.1,1.2) \) under the high-demand scenario, \( \alpha_i \sim U(0.8,0.9) \) for the low-demand scenario, and \( \alpha_i \sim U(0.8,1.2) \) for the mixed load scenario.

Each bus can assume one of the predefined load scaling factors for a system with \( n_d \) load buses. A fully exhaustive enumeration of all possible load combinations results in:
\vspace{-6pt}
\begin{equation}
    n_{\text{total}} = m^{n_d}
\end{equation}

where \(m\) is the number of discrete scaling values between 0.8 and 1.2. If continuous variation is considered, the number of scenarios would be infinite.  Hence, due to the exponential growth in scenarios, we need to use a method that realistically covers all possible load scenarios. 

To ensure \( \alpha_d \) varies across all load buses, we adopt LHS, a stratified sampling method that evenly partitions the input space and selects representative samples from each interval. Unlike random sampling, which may cluster points unevenly, LHS provides diverse and well-distributed load conditions \cite{iordanis2025efficacy}.

\subsection{Minimum Required Load Scenarios for Different System Sizes}

The minimum required number of load cases for even coverage in LHS is determined using the Monte Carlo error bound approach or the central limit theorem \cite{aistleitner2012central}:

\vspace{-6pt}
{\small \begin{equation}
    n_{\text{samples}} \geq \left( \frac{Z \sigma}{\epsilon} \right)^2
\end{equation}}
where \(Z\) is the critical value from the normal distribution for a 95\% confidence level, \(\sigma\) is the assumed normalized standard deviation (\(\approx 1\)), and \(\epsilon\) is the desired relative error tolerance.For a more conservative estimate in large systems, a common heuristic suggests setting the total number of samples as \(n_{\text{samples}} = 10 \times n_d\). 

The variance in OPF solutions is analyzed as a function of the number of load cases to validate that the generated scenarios are sufficient. If increasing \( n_{\text{samples}} \) beyond a certain point does not significantly change the OPF results, then additional cases are unnecessary. The mean squared deviation between OPF solutions is shown as:

\vspace{-8pt}
{\small \begin{equation}
    \sigma_X^2(n_{\text{samples}}) = \frac{1}{n_{\text{samples}}} \sum_{i=1}^{n_{\text{samples}}} || X_i^{\text{OPF}} - \bar{X}^{\text{OPF}} ||^2
\end{equation}}
where \( \bar{X}^{\text{OPF}} \) is the mean OPF solution across samples.

\begin{itemize}
    \item If \( \sigma_X^2 \) stabilizes with increasing \( N_{\text{samples}} \), the dataset is considered sufficient.
    \item If \( \sigma_X^2 \) continues to decrease, more samples are required.
\end{itemize}

Using these guidelines, Table~\ref{tableLoadScenarios} presents the estimated number of load scenarios. These values ensure robust GI-LSTM generalization across system sizes, improving both accuracy and computational efficiency.
\begin{table}[h]
\centering
\caption{Estimated Number of Load Scenarios}
\vspace{-6pt}
\label{tableLoadScenarios}
\begin{tabular}{lcc}
\toprule
System Size & \(n_d\) & \(n_{\text{samples}}\) \\
\midrule
Small  & \(n_d \leq 50\)             & 1000         \\
Medium & \(50 \leq n_d \leq 500\)    & 1000--2000   \\
Large  & \(n_d > 500\)               & 2000--10000  \\
\bottomrule
\end{tabular}
\end{table}
\vspace{-8pt}
\section{Numerical Results}
We evaluate the approach on a diverse set of test systems, ranging from small networks to large-scale, real-world cases, and examine its behavior under varying operating conditions. To provide insight into robustness, we consider a wide range of load scenarios and analyze feasibility/infeasibility detection and performance trends across these conditions. For comparison, we also report the convergence behavior of a classical warm-start prediction strategy, highlighting key differences in iteration dynamics and convergence behavior. 
\vspace{-6pt}
\subsection{Simulation Setting}
To evaluate the effectiveness and scalability of the proposed L-IPM, we conducted tests on power systems of varying sizes. Our analysis began with a small-scale 3-bus system and a standard 24-bus system, followed by larger systems comprising 118 and 2869 buses to assess the scalability of the method. The GI-LSTM model was implemented in Python using TensorFlow, which also enabled the implementation of the proposed grid-informed loss function by augmenting the standard prediction error with differentiable feasibility-penalty terms. The IPM-based OPF simulations were carried out using the MATPOWER 8.0 toolbox in MATLAB.

For the performance evaluation of the GI-LSTM model, we used multiple commonly used metrics, including mean squared error (MSE), mean absolute error (MAE), root mean squared error (RMSE), $R^2$ score, loss function trends, and residual analysis. We also compared the performance of the classic OPF and the L-IPM-based OPF using metrics, including the objective function value, feasibility condition error, complementarity condition error, and objective function error. These evaluations were conducted under various load scenarios, including adverse conditions with multiple active inequality constraints.
\vspace{-6pt}
\subsection{Look-Back Window Determination}
\begin{figure}[h]
    \centering
        \captionsetup{font={footnotesize}}
    \includegraphics[width=0.4\columnwidth]{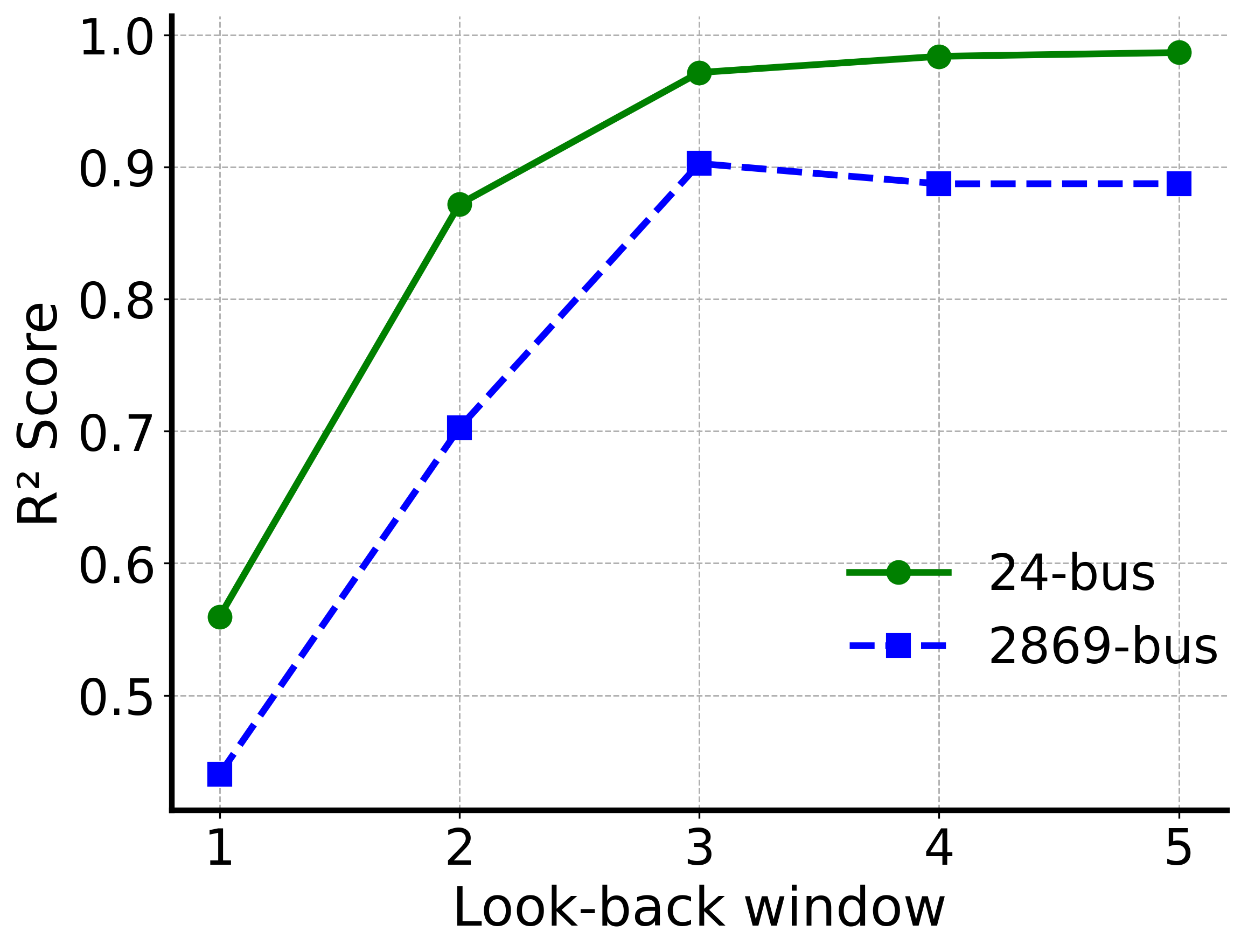}
    \vspace{-4pt}
    \caption{Comparison of R\textsuperscript{2} scores for different look-back values.}
    \label{fig:lookback}
\end{figure}

To determine the appropriate number of early iterations to input into the GI-LSTM, we conducted multiple simulations. A single iteration does not provide enough historical context for the GI-LSTM to capture temporal patterns and interdependencies. Therefore, we increased the number of early iterations, or the look-back window, from one to five and observed that in many cases, the output became more accurate with a higher number of iterations. However, there were instances where increasing the number of iterations led to a slight decrease in accuracy. Fig.~\ref{fig:lookback} reports the \(R^2\) scores for two representative systems for clarity. We observe that setting the look-back to three provides satisfactory performance, and for the 2869-bus system, it achieves the best performance overall. Moreover, in many cases, increasing the look-back from three to four or five does not significantly improve precision. Based on these observations, we selected a look-back of three iterations to feed into the GI-LSTM.

\vspace{-14pt}

\subsection{GI-LSTM Performance}
\begin{figure}[h]
    \centering
        \captionsetup{font={footnotesize}}
    \subfloat[3-bus system]{\includegraphics[width=0.4\columnwidth]{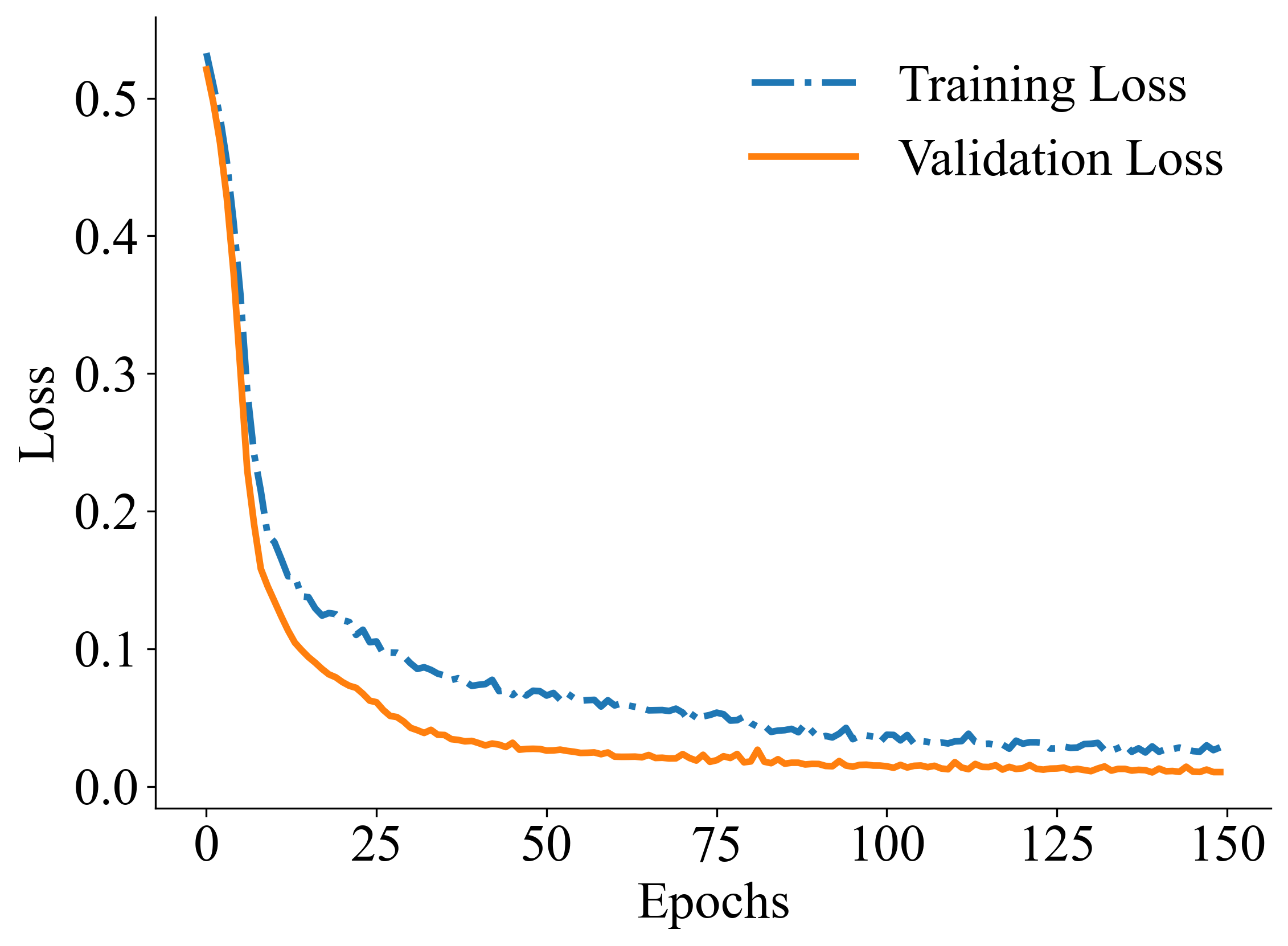}}
    \hfill
    \subfloat[2869-bus system]{\includegraphics[width=0.4\columnwidth]{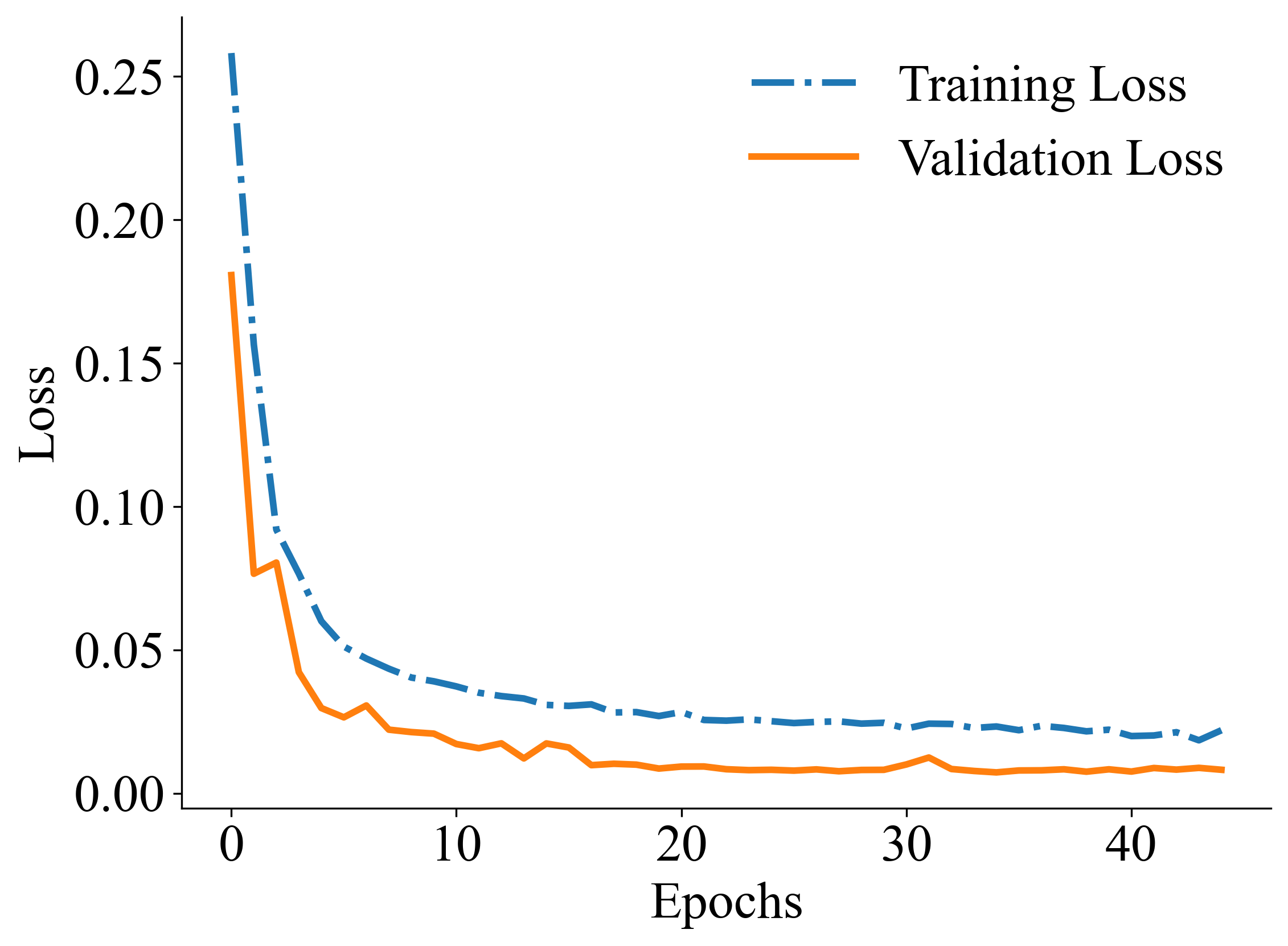}}
    \caption{GI-LSTM loss function error.}
    \vspace{-4pt}
    \label{lossfunc}
\end{figure}

The evaluation metrics of GI-LSTM demonstrate its effectiveness. For instance, the loss function of the trained GI-LSTM on two test systems is shown in Fig.~\ref{lossfunc}. We observed that both the training and validation losses converged toward small values, indicating that the GI-LSTM model was effectively optimized and capable of learning meaningful patterns from the data. In this model, a slightly lower validation loss is expected compared to the training loss. This primarily results from dropout, which is active during training but disabled during validation, making training inherently noisier because random units are temporarily zeroed at each iteration. Consequently, training is performed under a deliberately perturbed network, whereas validation uses the full model capacity. In addition, training loss is averaged over mini-batches during weight updates, whereas validation loss is computed after each epoch using the updated parameters. Overall, this behavior is consistent with stable regularized training and improved generalization.

A high coefficient of determination $R^2$ across all four test systems further confirms the model's predictive accuracy, as it demonstrates a strong alignment between the predicted and actual values in the validation phase. These high-accuracy projections are verified with IPM. The performance metrics for each system are summarized in Table~\ref{tab:R2Scores}, highlighting the consistency and generalizability of the proposed approach across different system sizes.

\vspace{-8pt}
\begin{table}[h]
\centering
\vspace{-6pt}
\caption{R\textsuperscript{2} Score for GI-LSTM Across Different Systems}
\vspace{-4pt}
\label{tab:R2Scores}
\begin{tabular}{lcccc}
\toprule
         & 3-bus & 24-bus & 118-bus & 2869-bus \\
\midrule
ACOPF    & 0.981 & 0.977  & 0.994   & 0.902    \\
DCOPF    & 0.990 & 0.971  & 0.983   & 0.949    \\
\bottomrule
\end{tabular}
\end{table}
\vspace{-8pt}
In Table~\ref{tab:R2Scores}, we observe that the R\textsuperscript{2} score for ACOPF in the 2869-bus system drops to 0.9. This decline is primarily attributed to the nature of the GI-LSTM model, which attempts to enforce all solutions within the feasible region. However, we later observed that not all demand scenarios have feasible solutions when solved using IPM. Therefore, the observed drop is largely because the solutions remain constrained within the feasible region. Table~\ref{tableMSE} shows additional metrics used to evaluate the trained GI-LSTM model, further supporting our approach's strong performance and reliability. 

\begin{table}[h]
\centering
\caption{GI-LSTM Prediction Errors Across Different Systems}
\vspace{-5pt}
\label{tableMSE}
\begin{tabular}{lcccc}
\toprule
      & 3-bus & 24-bus & 118-bus & 2869-bus \\
\midrule
MSE   & 0.010 & 0.012  & 0.004   & 0.080    \\
MAE   & 0.068 & 0.067  & 0.033   & 0.120    \\
RMSE  & 0.100 & 0.110  & 0.068   & 0.280    \\
\bottomrule
\end{tabular}
\end{table}

\vspace{-12pt}
\subsection{GI-LSTM Outpout}
\begin{figure}[!t]
    \centering
        \captionsetup{font={footnotesize}}
    \subfloat[ACOPF]{\includegraphics[width=0.4\columnwidth]{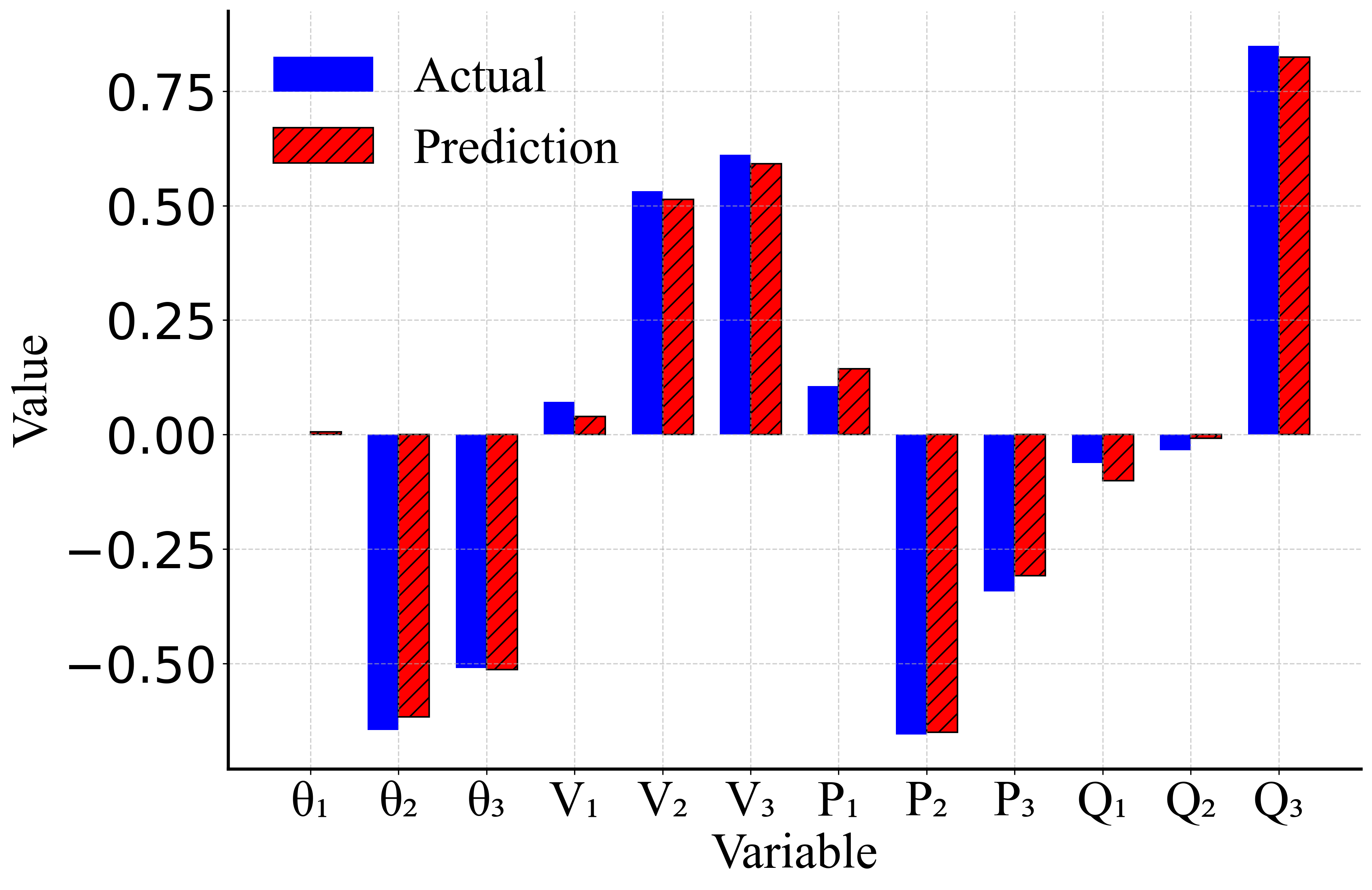}}
    \hfill
    \subfloat[DCOPF]{\includegraphics[width=0.4\columnwidth]{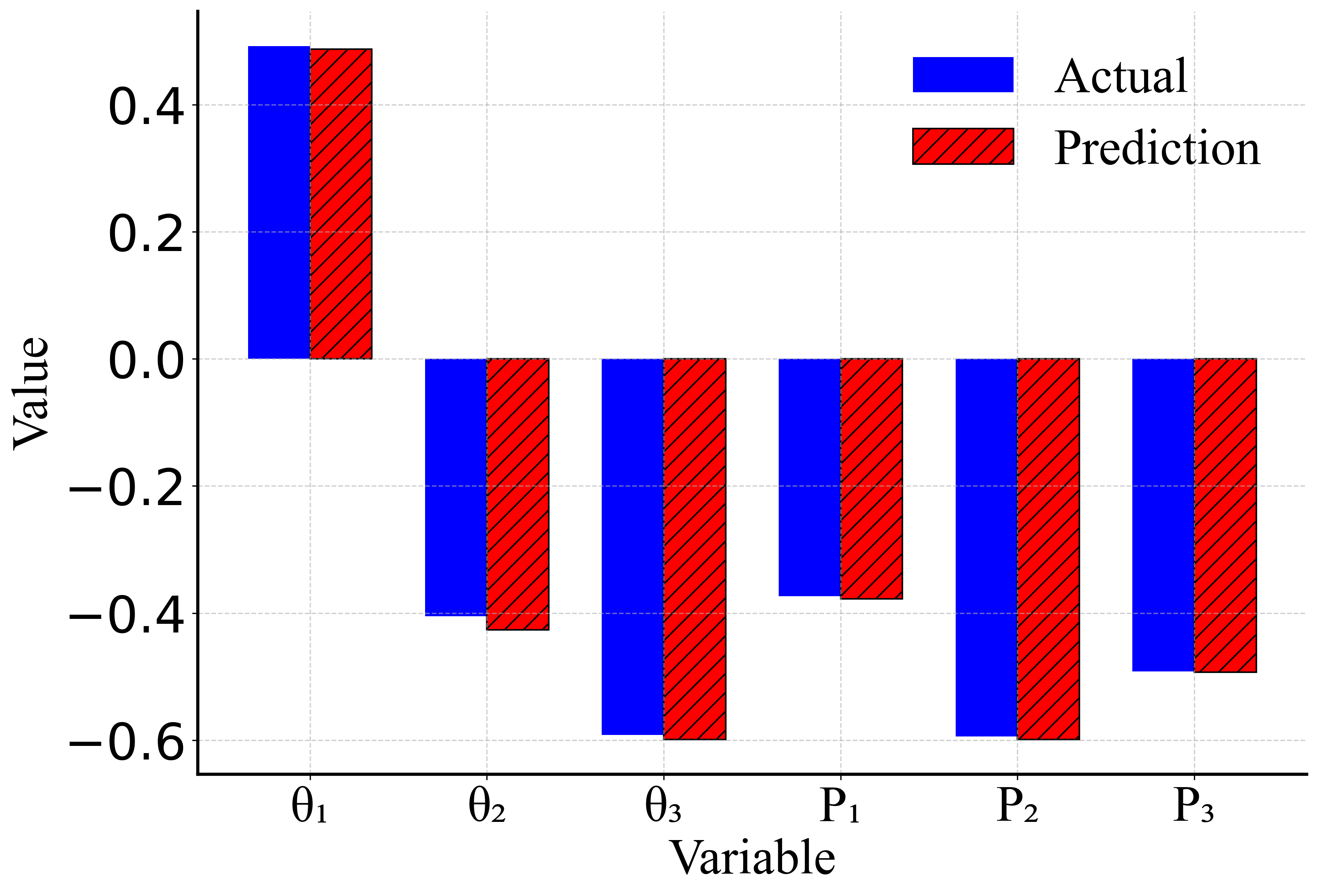}}
    \caption{GI-LSTM output comparison with IPM in the 3-bus system.}
    \label{LSTMoutput}
\end{figure}

The GI-LSTM model predicts outputs that are both accurate and close to the exact solution. For instance, as Fig.~\ref{LSTMoutput}, the results generated by GI-LSTM closely match those obtained using IPM, which demonstrates the model’s strong predictive capability. Even minor violations of operational limits—such as voltage, power flow, or generation constraints—can pose serious risks to system stability and reliability. To address this, we sanity check the GI-LSTM predictions by implementing IPM using the GI-LSTM projected central path. This step ensures that the final solution remains the same as the optimal OPF solution and strictly satisfies all system physical and operational constraints. Hence, the proposed GI-LSTM promotes high-quality L-IPM OPF solutions by steering the trajectory toward the feasible region, while a final IPM step verifies solution accuracy.

\subsection{L-IPM}

\begin{figure}[!t]
    \centering
        \captionsetup{font={footnotesize}}
    \subfloat[Running Time]{%
        \includegraphics[width=0.4\columnwidth]{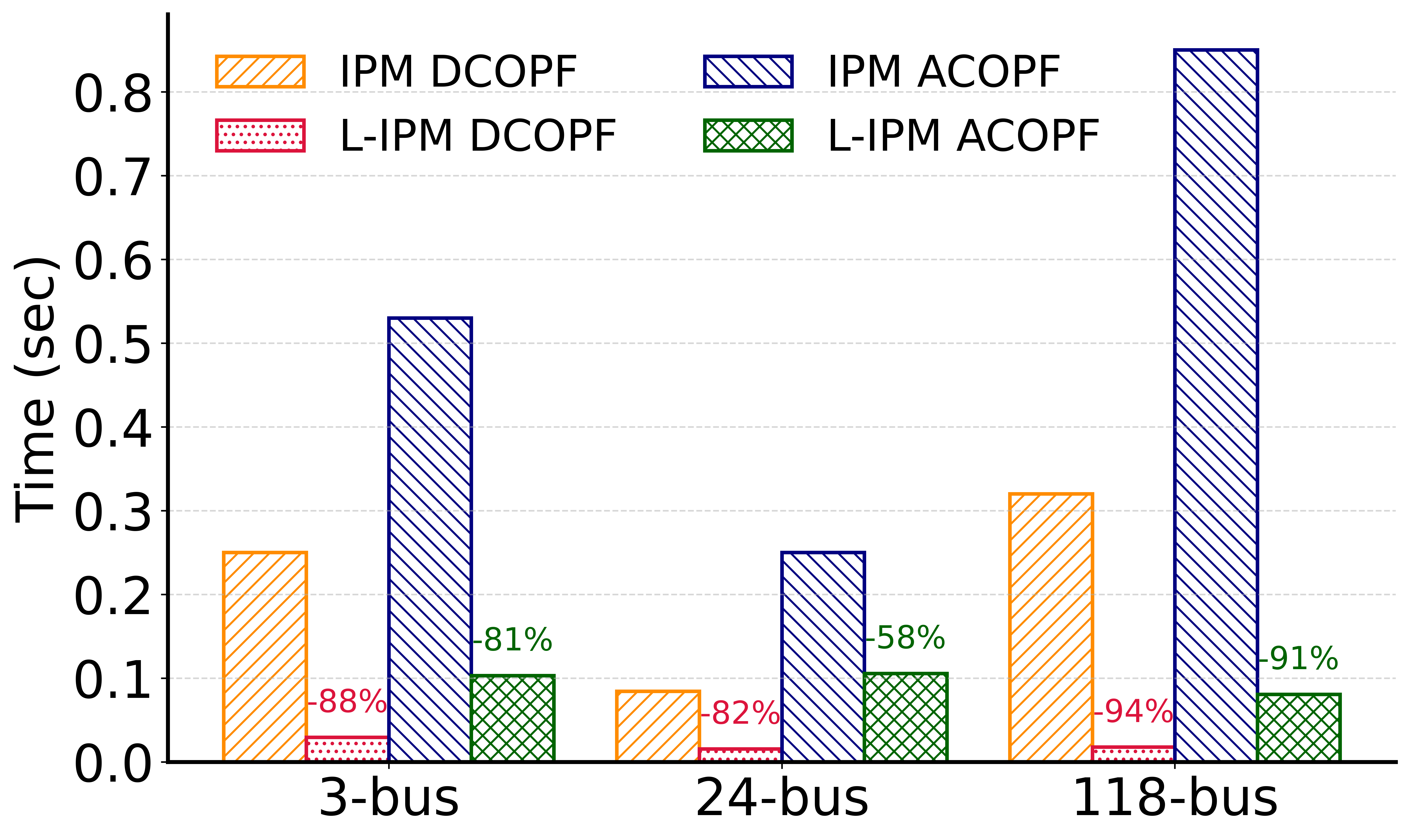}%
    }
    \hfill
    \subfloat[Running Time]{%
        \includegraphics[width=0.38\columnwidth]{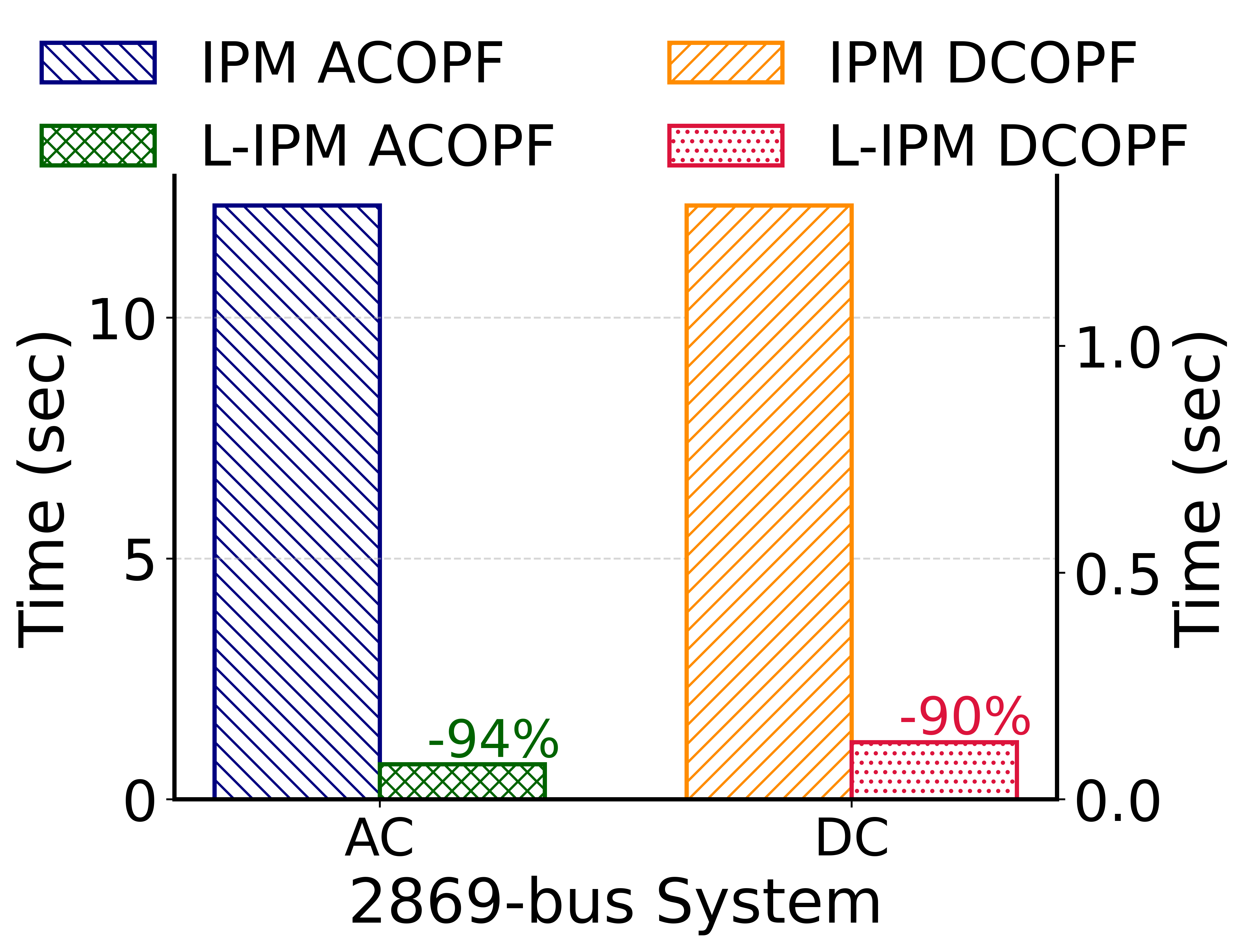}%
    }
    \vspace{0.8em}
    \subfloat[Iteration Comparison]{%
        \includegraphics[width=0.4\columnwidth]{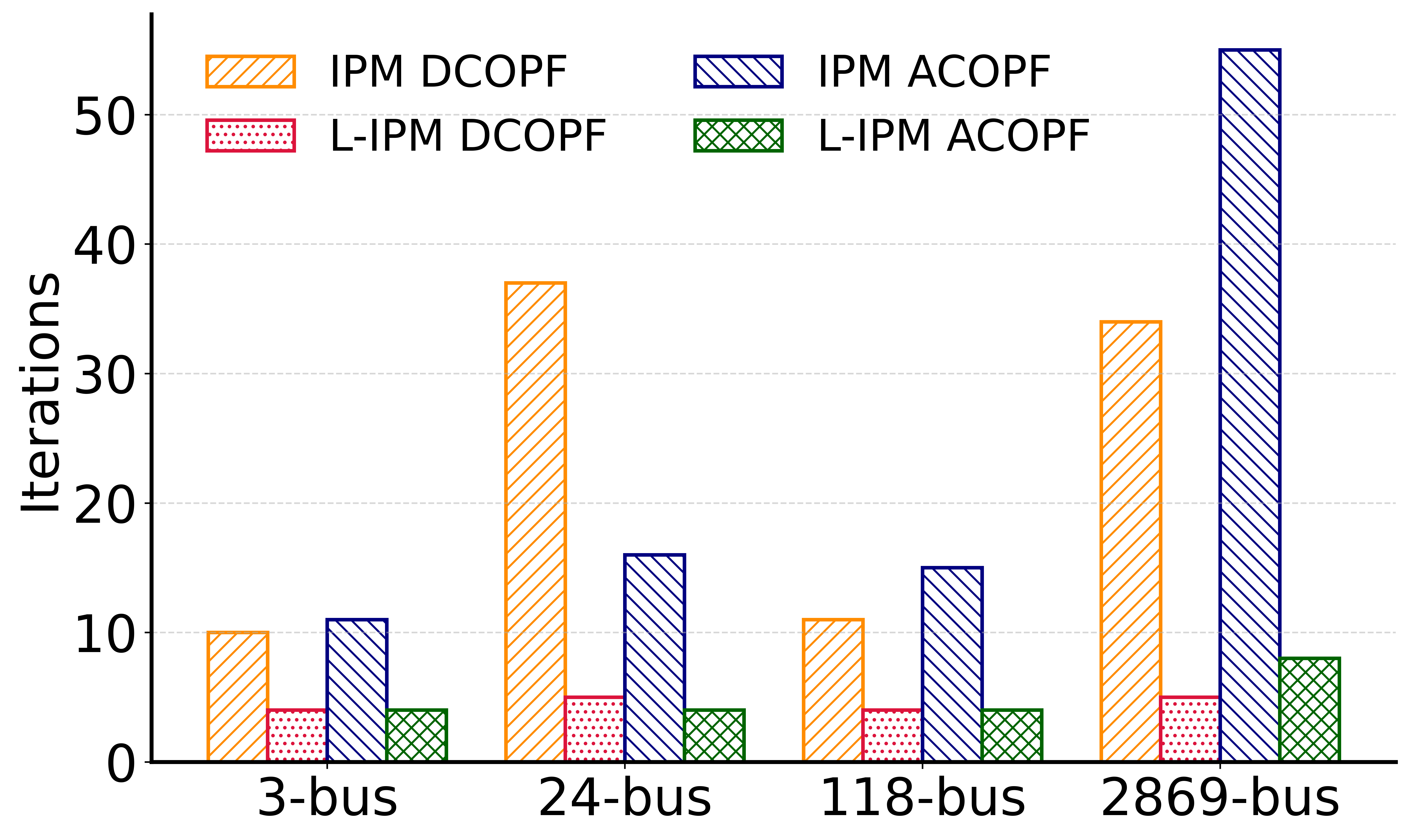}%
    }
    \caption{Performance comparison of IPM and L-IPM.}
    \label{fig:performanceComp}
\end{figure}

\emph{Simulation Results Summary:} 
Table~\ref{tableIter} reports the minimum and maximum number of iterations required to solve the OPF problem across all load scenarios for each test system. As system size increases, the classical IPM becomes more sensitive to load variations, leading to a larger gap between the minimum and maximum iteration counts, mainly due to more active inequality constraints under stressed conditions. In contrast, the number of L-IPM iterations shows no such variation, highlighting its robustness to load fluctuations. The percentage reduction in average iterations of L-IPM relative to classical IPM is shown in Fig.~\ref{fig:performanceComp}(c); for example, in ACOPF on the 2869-bus system, L-IPM reduces the average number of iterations by 85.5\%, representing a substantial improvement.

\begin{table}[t]
\centering
\caption{OPF Speedup Comparison of L-IPM over IPM\textsuperscript{\dag}}
\vspace{-5pt}
\label{tableIter}
\setlength{\tabcolsep}{4pt}      
\renewcommand{\arraystretch}{1.05}
\footnotesize
\begin{tabular}{lcc|cc}
\toprule
\multirow{2}{*}{} & \multicolumn{2}{c|}{Avg. Speedup ($\times$)} & \multicolumn{2}{c}{Enhancement (\%)} \\
                  & DC & AC & min & max \\
\midrule
3-bus     & 2.50 & 2.88 & \cellcolor{gray!20}\textbf{60}   & \cellcolor{gray!20}\textbf{66.7} \\
24-bus    & 4.70 & 3.63 & \cellcolor{gray!20}\textbf{50}   & \cellcolor{gray!20}\textbf{86.5} \\
118-bus   & 2.50 & 3.63 & \cellcolor{gray!20}\textbf{55.6} & \cellcolor{gray!20}\textbf{73.4} \\
2869-bus  & 4.78 & 5.22 & \cellcolor{gray!20}\textbf{63.7} & \cellcolor{gray!20}\textbf{85.5} \\
\bottomrule
\end{tabular}

\vspace{0.5mm}
\begin{flushleft}
\footnotesize
\vspace{-6pt}
\textsuperscript{\dag}Avg. speedup ($\times$) is computed as the average speedup over the analyzed OPF cases for each OPF type; values $>1$ indicate that L-IPM is faster.
\end{flushleft}
\end{table}

The iterations avoided by the proposed L-IPM correspond to the most time-consuming and computationally challenging phases of the classic IPM. These later iterations are typically associated with ill-conditioned systems that are computationally expensive to solve, as shown in Fig.~\ref{fig:time_cond}(a). The observed reduction in runtime exceeds the reduction in iteration count. Table~\ref{tableRuntime} reports the total runtime for different test cases under worst-case demand scenarios. It compares the performance of IPM and L-IPM for both AC and DCOPF. As seen, L-IPM achieves substantial runtime savings across all test cases. Fig.~\ref{fig:performanceComp}(a) and (b) visualize the computational time comparison between IPM- and L-IPM-based OPF solutions, performed on a machine with 16~GB RAM and an Intel Core i9 processor. Given the different solution time scales, and for improved visualization, we present the results for the 2869-bus system in a separate figure. A significant reduction in runtime is observed across all cases for both ACOPF and DCOPF problems, particularly for larger, real-world-scale systems. For example, in the case of ACOPF on the 2869-bus European high-voltage transmission system in Fig.~\ref{fig:performanceComp}(b), the proposed L-IPM-based approach is 94\% faster than the classical IPM-based OPF in finding the same solution as classical IPM.
\begin{table}[h]
\centering
\caption{OPF Runtime Comparison of IPM and L-IPM}
\vspace{-6pt}
\label{tableRuntime}
\begin{tabular}{lcc|cc|cc}
\toprule
\multirow{2}{*}{} & \multicolumn{2}{c|}{IPM (ms)} & \multicolumn{2}{c|}{L-IPM (ms)} & \multicolumn{2}{c}{Time Reduction (\%)} \\
                  & AC & DC                  & AC & DC                  & AC & DC \\
\midrule
3-bus     & 530   & 250   & 103 & 29  & \cellcolor{gray!20}\textbf{80.6} & \cellcolor{gray!20}\textbf{88.3} \\
24-bus    & 250   & 84    & 105 & 15  & \cellcolor{gray!20}\textbf{57.8} & \cellcolor{gray!20}\textbf{81.6} \\
118-bus   & 850   & 320   & 80  & 17  & \cellcolor{gray!20}\textbf{90.5} & \cellcolor{gray!20}\textbf{94.5} \\
2869-bus  & 12330 & 1310  & 730 & 126 & \cellcolor{gray!20}\textbf{94.1} & \cellcolor{gray!20}\textbf{90} \\
\bottomrule
\end{tabular}
\end{table}

Below, we present a detailed analysis of each test system to provide further insights into the simulation results.

\subsubsection{3-bus system}
\begin{figure}[!t]
    \centering
        \captionsetup{font={footnotesize}}
    \subfloat[IPM ACOPF]{\includegraphics[width=0.4\columnwidth]{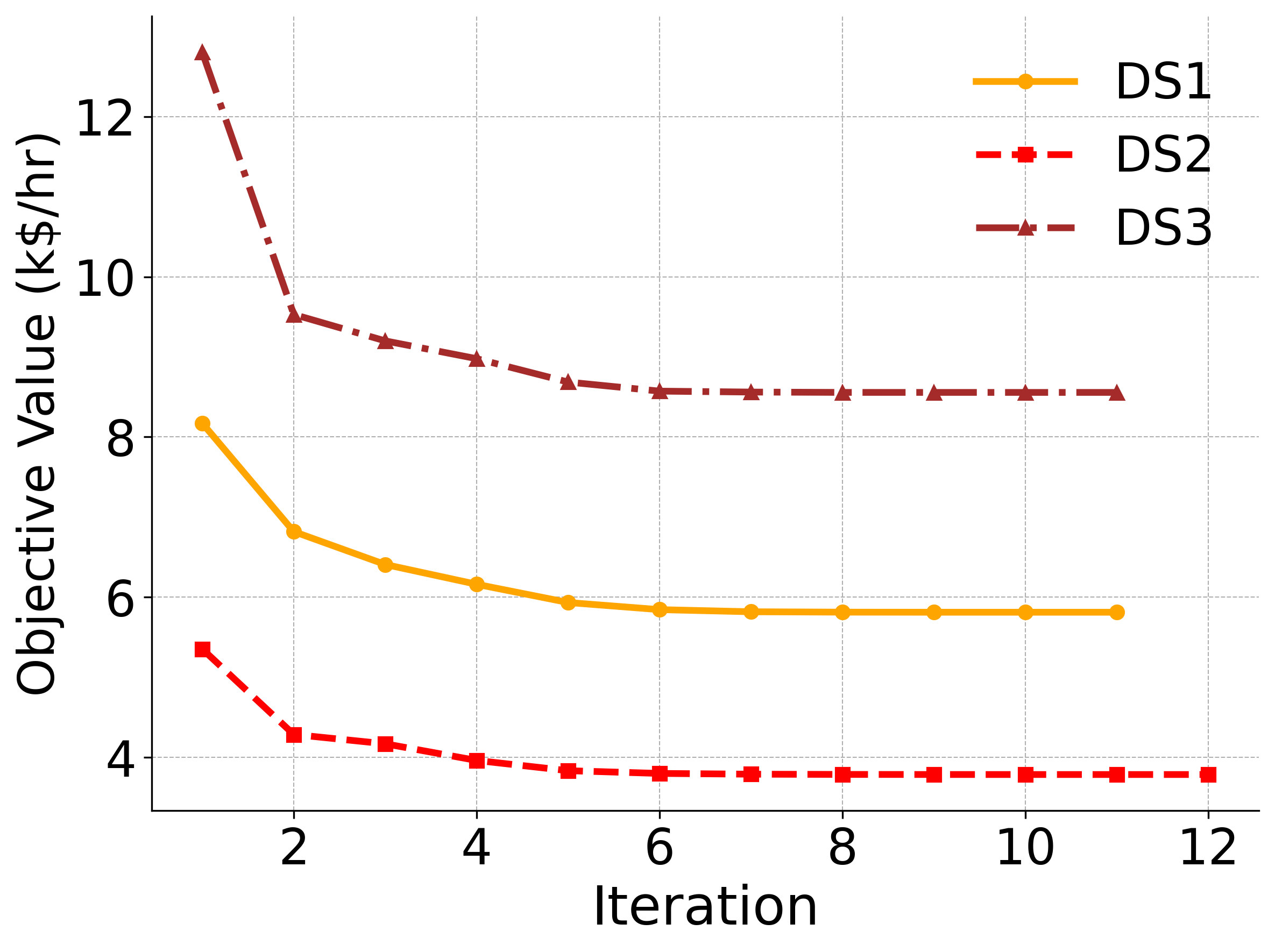}}\hfill
    \subfloat[L-IPM ACOPF]{\includegraphics[width=0.4\columnwidth]{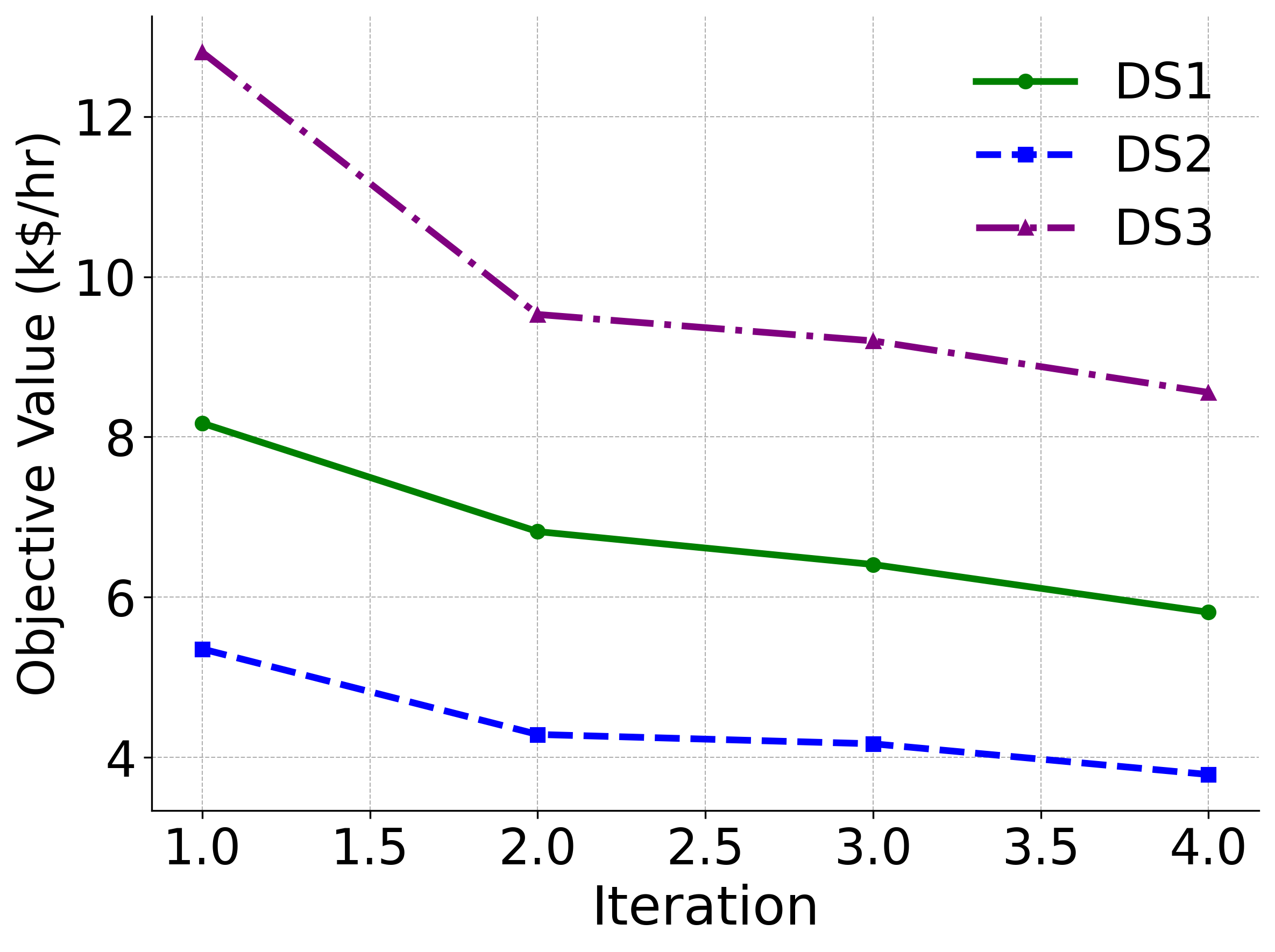}}\\
    \subfloat[IPM DCOPF]{\includegraphics[width=0.4\columnwidth]{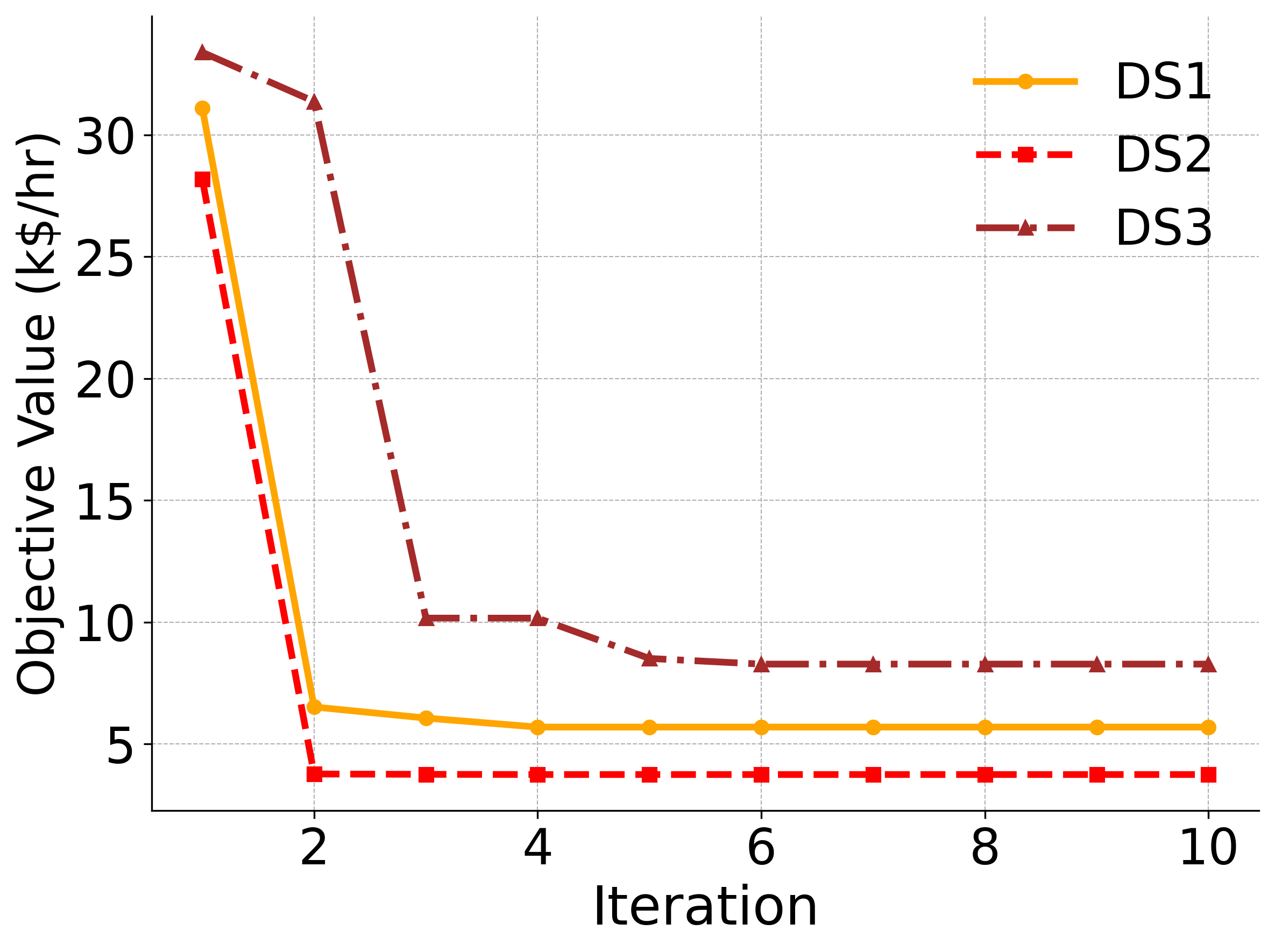}}\hfill
    \subfloat[L-IPM DCOPF]{\includegraphics[width=0.4\columnwidth]{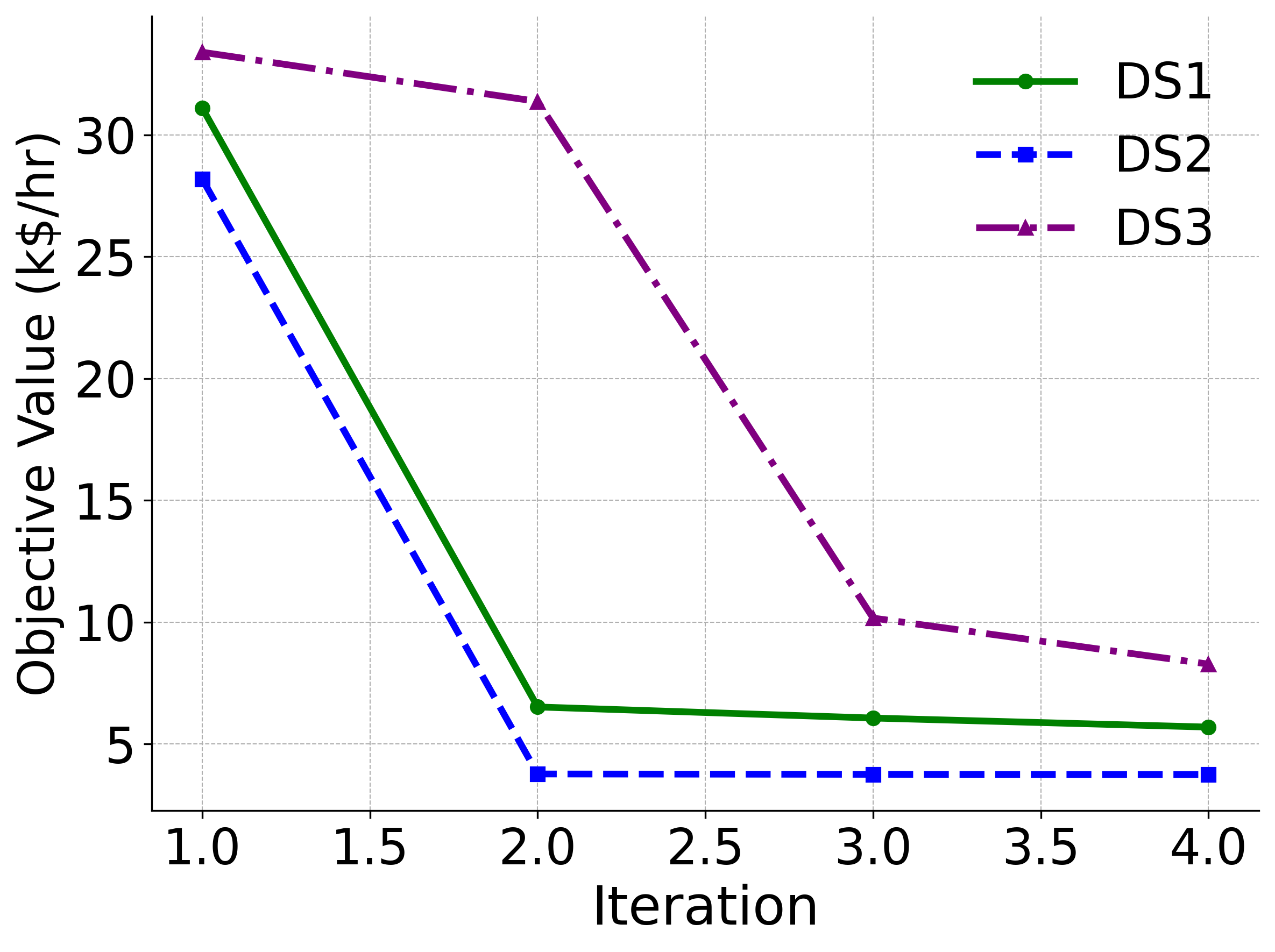}}
    \caption{OPF objective function values for the 3-bus system.}
    \label{fig:3_obj}
\end{figure}

We report detailed results for three demand scenarios (DSs): low load, medium load, and high load. In Figs. \ref{fig:3_obj}(a) and (c), the classic IPM is used to solve the DSs, and the OPF objective function values are reported across the iterations. Figs.~\ref{fig:3_obj}(b) and (d) show the performance of L-IPM, where the GI-LSTM output, trained based on observing just the first three iterations, is refined using IPM. L-IPM reaches the same solution as IPM but with fewer iterations, highlighting how L-IPM helps accelerate convergence while maintaining accuracy.

\begin{figure}[!t]
    \centering
        \captionsetup{font={footnotesize}}
    \includegraphics[width=0.4\columnwidth]{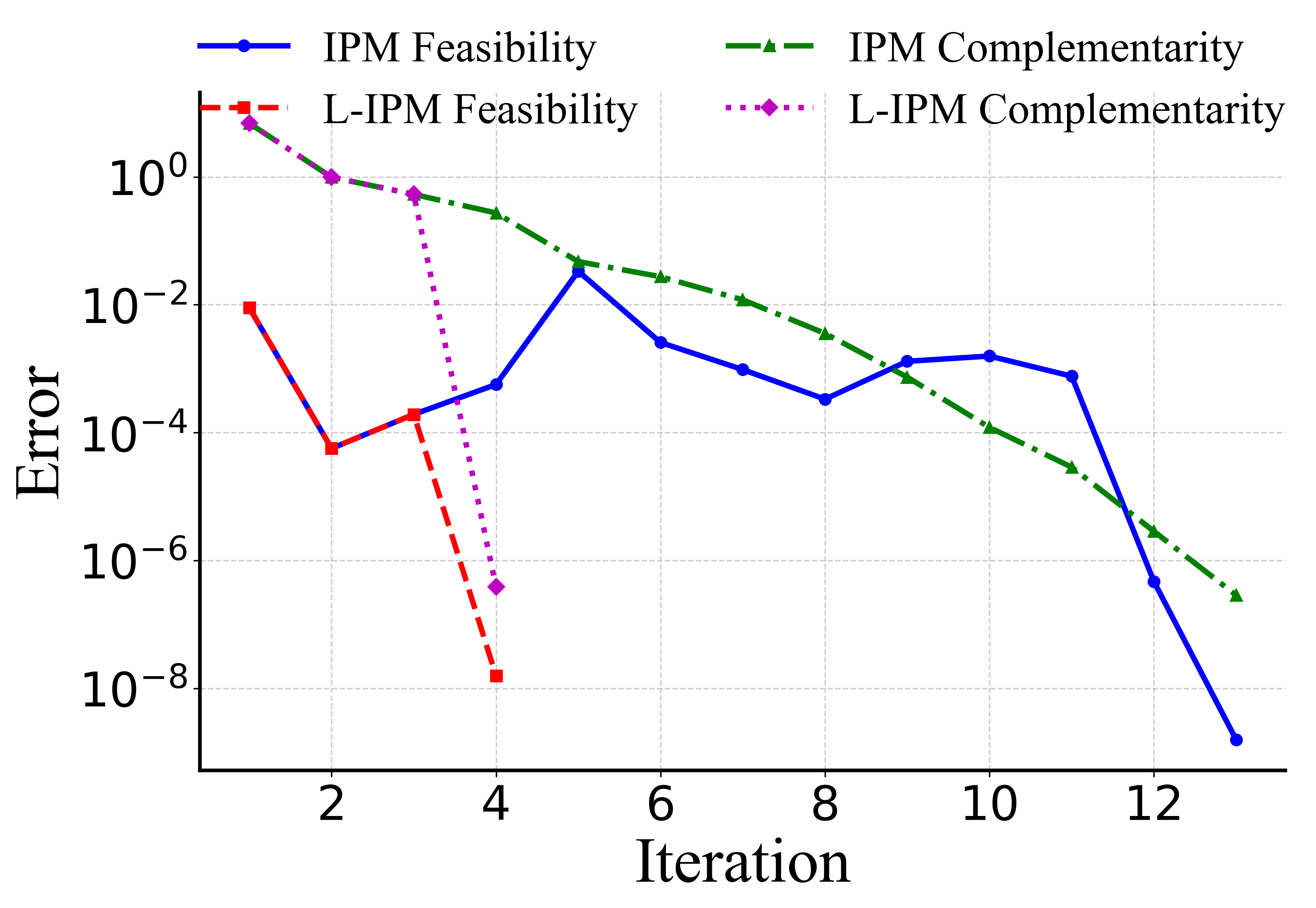}
    \caption{Feasibility and complementarity condition comparison between  IPM and L-IPM in 3-bus system.}
        \vspace{-10pt}
    \label{3FeasCom}
\end{figure}

For a more detailed comparison, we evaluated the classical IPM and the proposed L-IPM feasibility and complementarity conditions. As Fig.~\ref{3FeasCom}, L-IPM significantly reduces the error to below \(10e-6\). We analyzed the structure of the Newton steps, particularly the \(A\) matrix in the linear system \(Ax = b\), and observed that most of the significant changes occur during the early iterations. The remaining iterations primarily refine the solution by accurately identifying violated constraints and adjusting the slack variables within the IPM framework. The L-IPM learns the behavior of these variables, and with a carefully selected set of load scenarios, as described in the data generation section, it develops an understanding of the feasible region of the system, including generation limits and line capacities.

In the 3-bus system case, we also observed that most solution trajectories were similar. The main difference was in the objective function value. This consistent and well-behaved nature of IPM helped L-IPM accurately capture and predict solutions that are very close to the exact optimal values.

\subsubsection{24-bus system}
For this system, we observed that a higher number of IPM iterations was often required to reach a solution. In particular, under extreme load scenarios, the solver needed more than 30 iterations in many cases. 

\begin{figure}[!t]
    \centering
        \captionsetup{font={footnotesize}}
    \subfloat[IPM ACOPF]{\includegraphics[width=0.4\columnwidth]{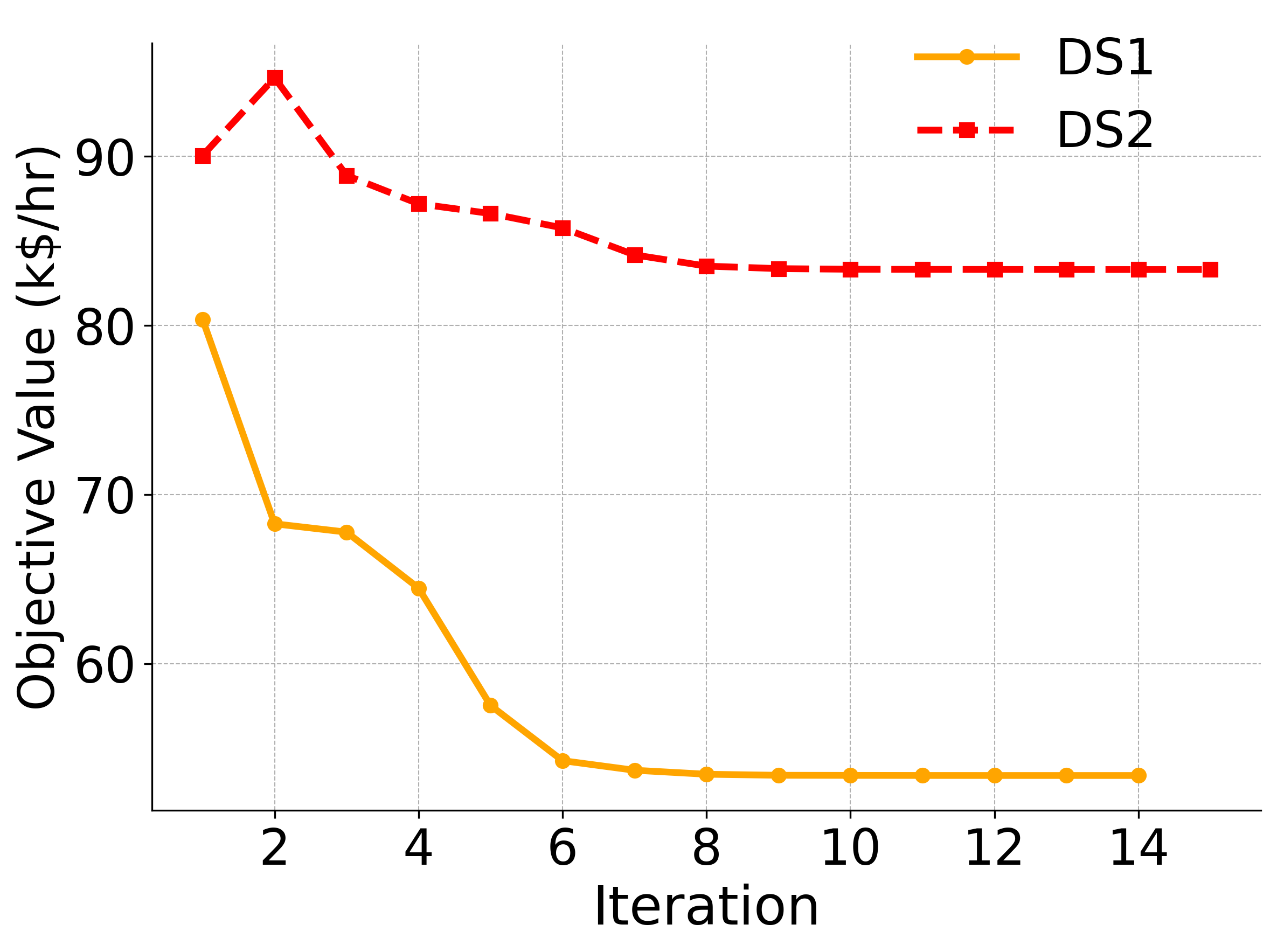}}\hfill
    \subfloat[L-IPM ACOPF]{\includegraphics[width=0.4\columnwidth]{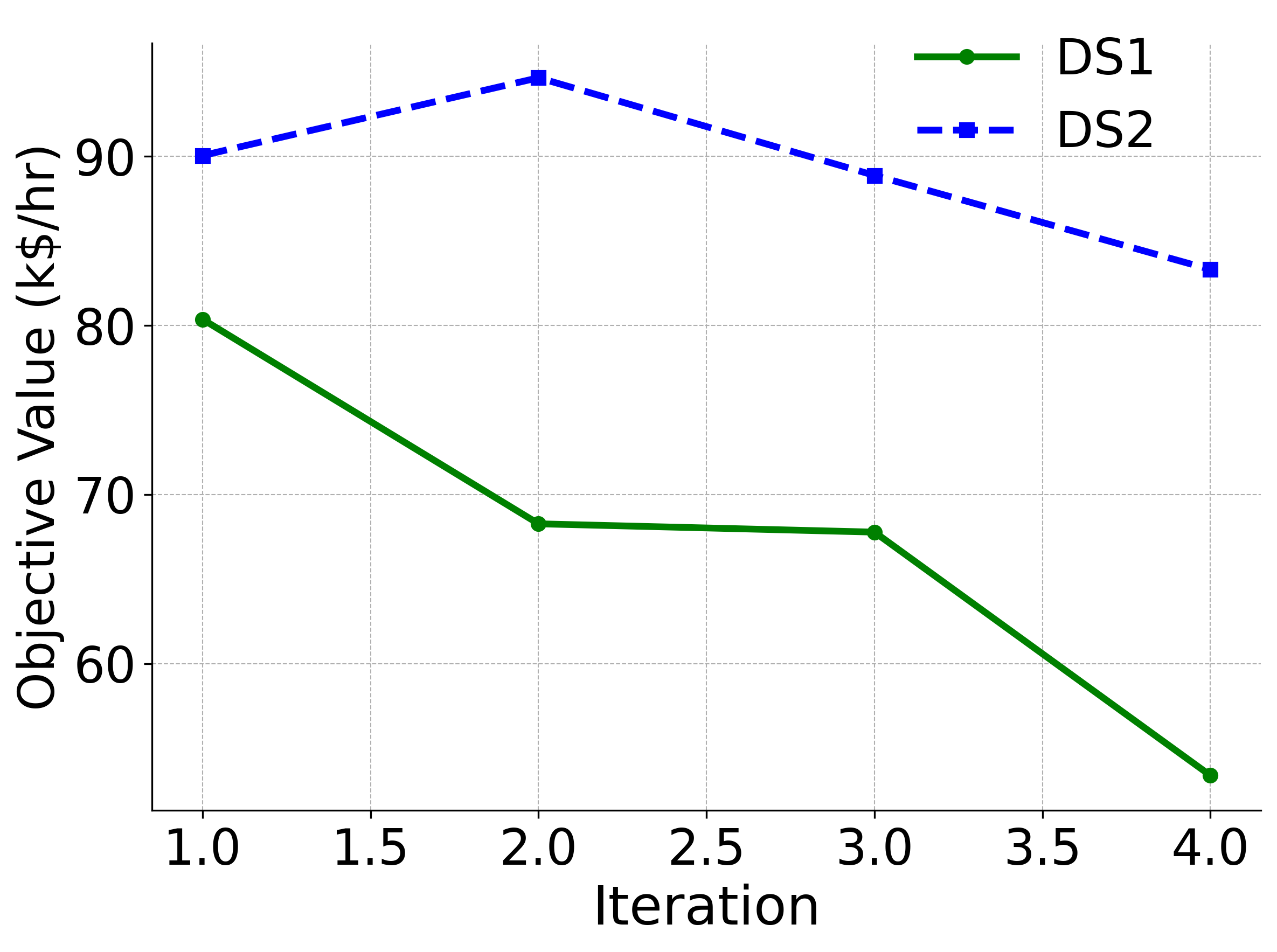}}\\
    \subfloat[IPM DCOPF]{\includegraphics[width=0.4\columnwidth]{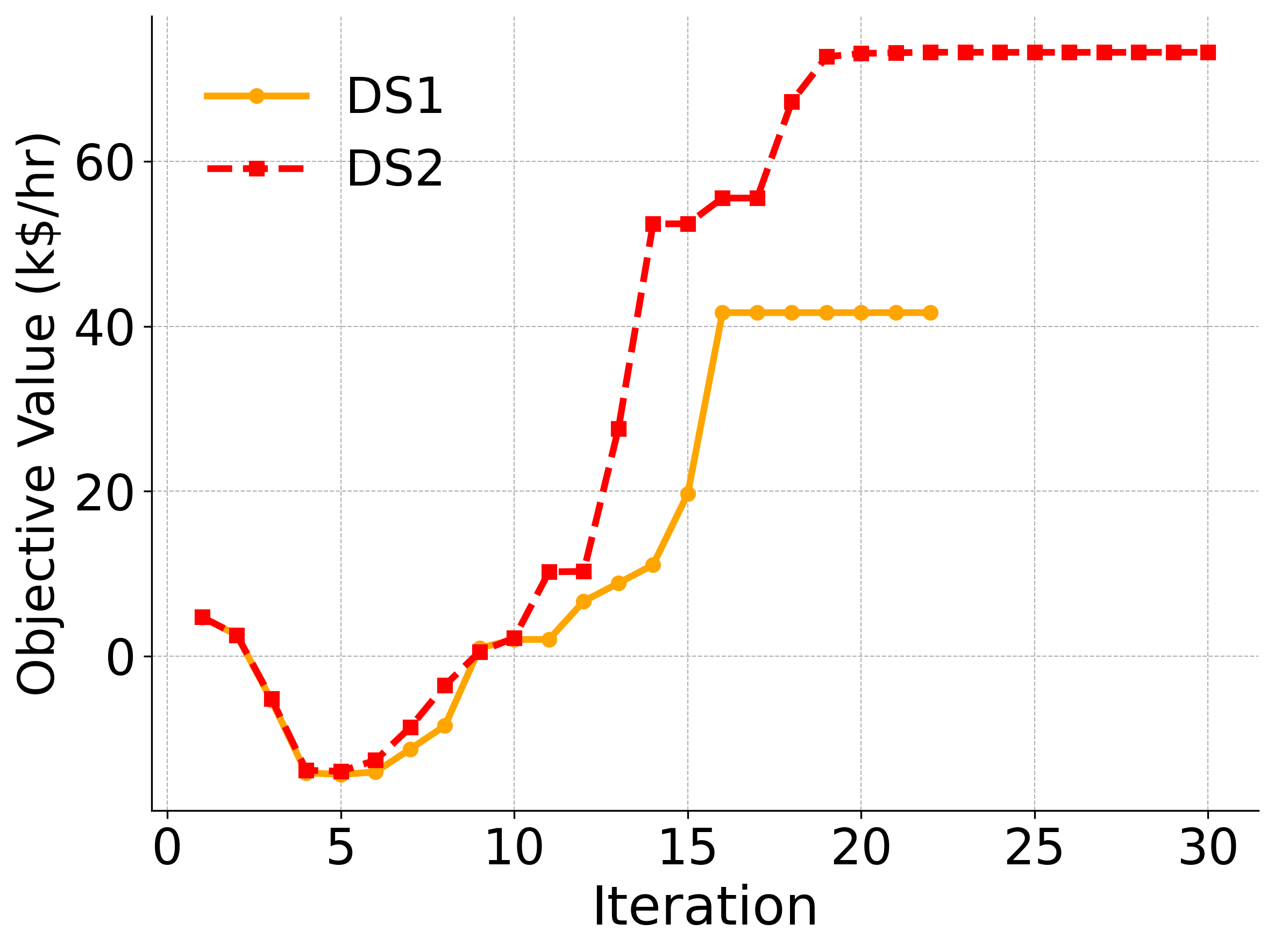}}\hfill
    \subfloat[L-IPM DCOPF]{\includegraphics[width=0.4\columnwidth]{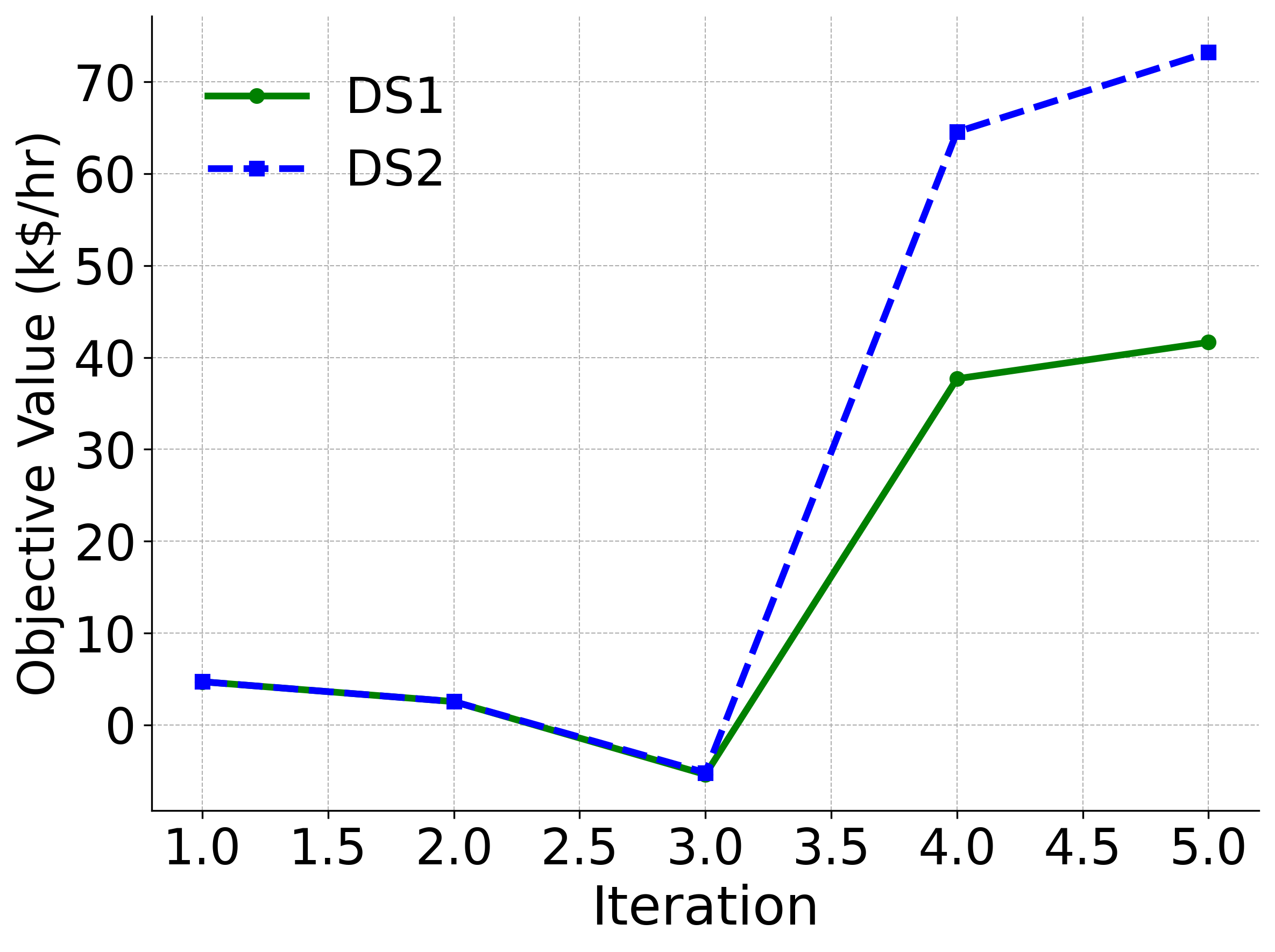}}
    \caption{OPF objective function values for the 24-bus system.}
        \vspace{-10pt}
    \label{fig:24_obj}
\end{figure}

As illustrated in Fig.~\ref{fig:24_obj}, we compare the performance of ACOPF and DCOPF for two representative cases. The first case involves a load scenario close to the nominal operating point, while the second represents an extreme load condition. The figure illustrates the increased number of iterations required by IPM to solve this load scenario. For both load scenarios, L-IPM converges in just four iterations for ACOPF and five iterations for DCOPF.

\begin{figure}[h]
    \centering
        \captionsetup{font={footnotesize}}
    \subfloat[ACOPF]{\includegraphics[width=0.4\columnwidth]{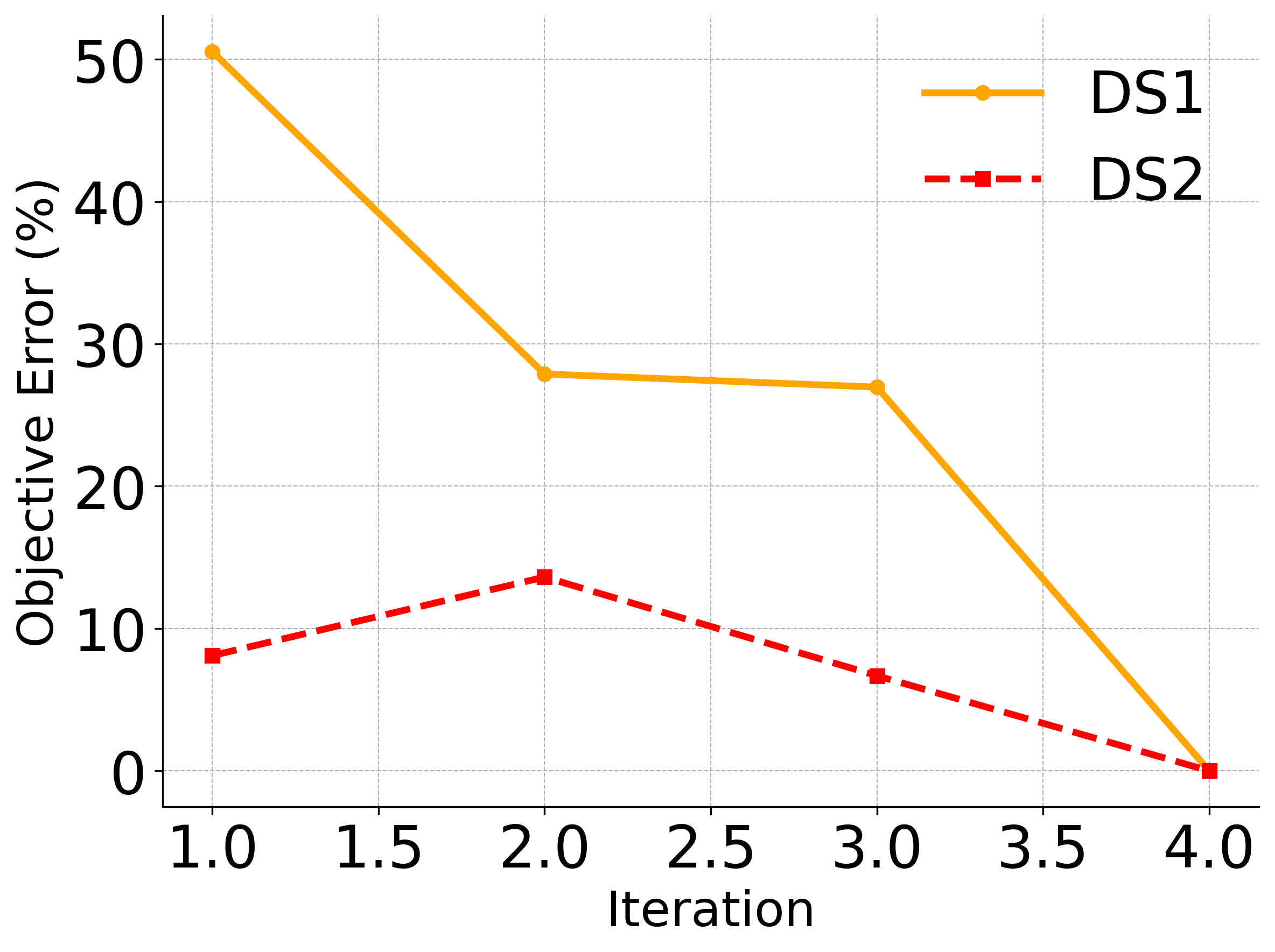}}
    \hfill
    \subfloat[DCOPF]{\includegraphics[width=0.4\columnwidth]{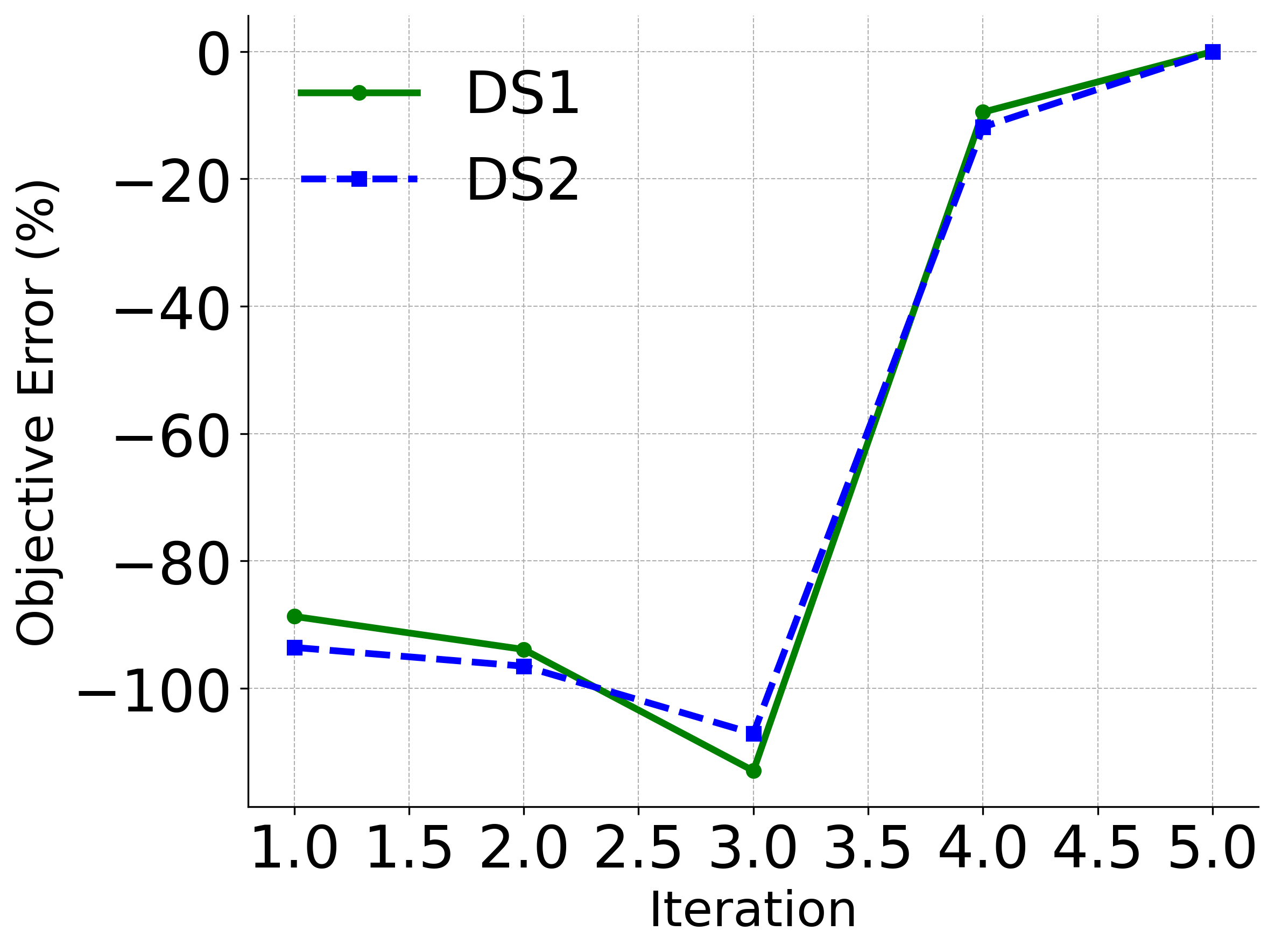}}
    \caption{L-IPM OPF objective function error of the 24-bus system.}
        \vspace{-10pt}
    \label{24LSTMerror}
\end{figure}

\begin{figure}[h]
    \centering
        \captionsetup{font={footnotesize}}
    \includegraphics[width=0.4\columnwidth]{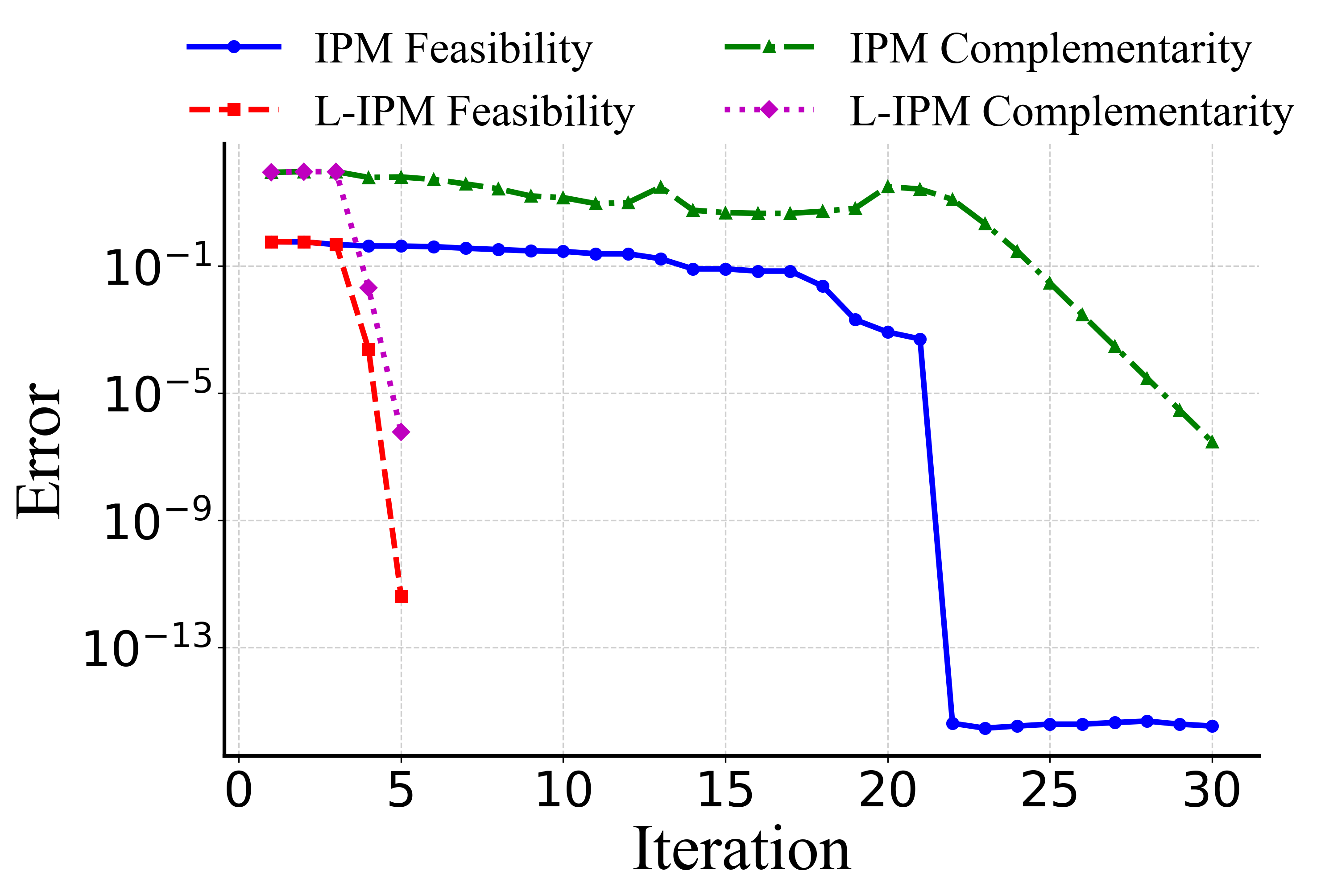}
    \vspace{-4pt}
    \caption{Feasibility and complementarity condition comparison between  IPM and L-IPM in 24-bus system.}
        \vspace{-10pt}
    \label{24FeasCom}
\end{figure}

To better illustrate how efficiently L-IPM reduces the objective function relative error compared to IPM, Fig.~\ref{24LSTMerror} presents the relative error of L-IPM for both DCOPF and ACOPF. Fig.~\ref{24FeasCom} presents a comparison of feasibility and complementarity condition errors under an extreme load scenario, further highlighting the robustness and efficiency of L-IPM. 


\subsubsection{118-bus system}
We selected this system to explore whether larger systems tend to exhibit similar convergence patterns as seen in the 24-bus system. The relatively stable behavior observed in the 118-bus case suggests that convergence characteristics may vary depending on the specific structure and properties of each power system, rather than system size alone.

\begin{figure}[!t]
    \centering
        \captionsetup{font={footnotesize}}
    \subfloat[IPM ACOPF]{\includegraphics[width=0.4\columnwidth]{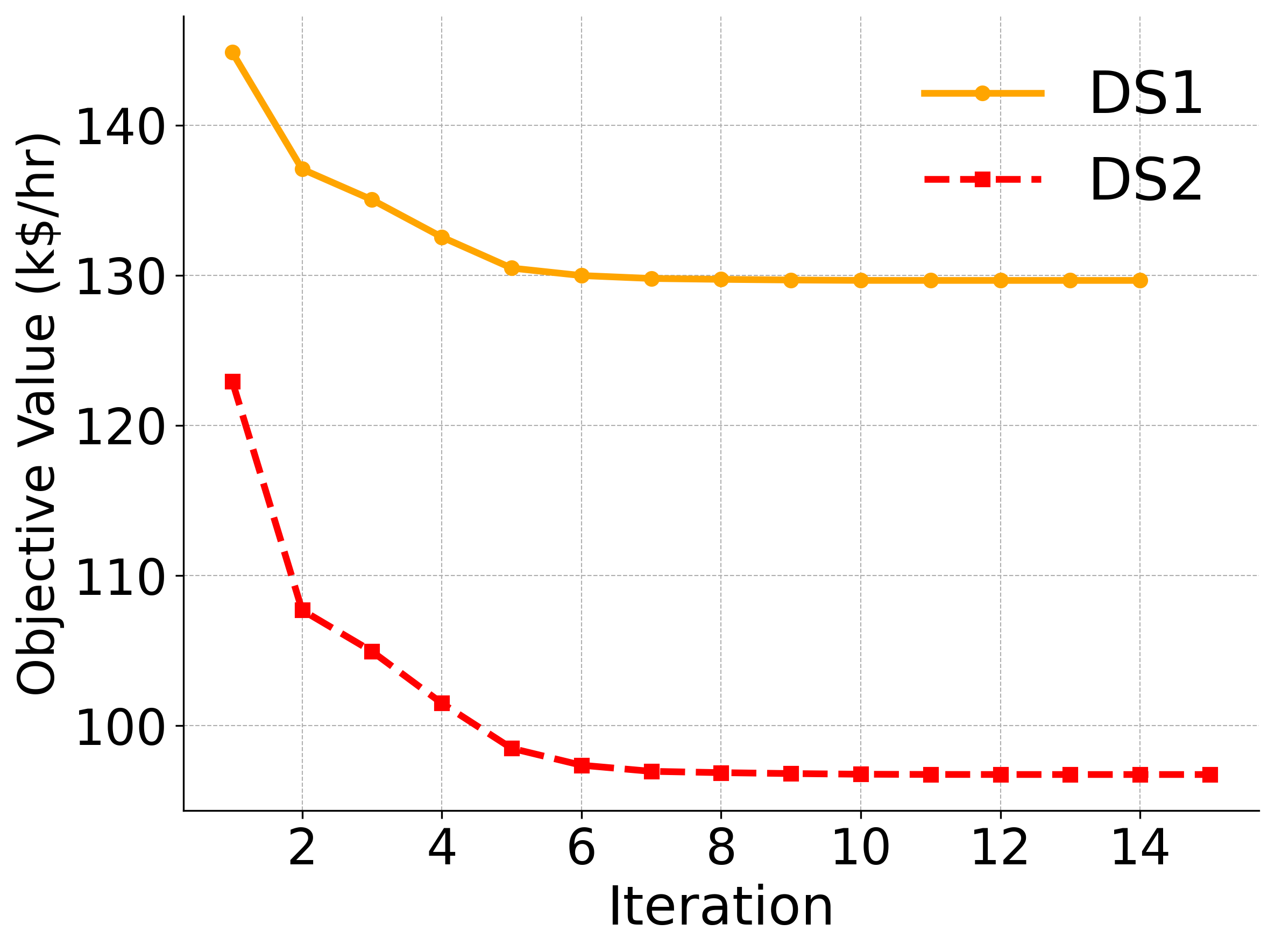}}\hfill
    \subfloat[L-IPM ACOPF]{\includegraphics[width=0.4\columnwidth]{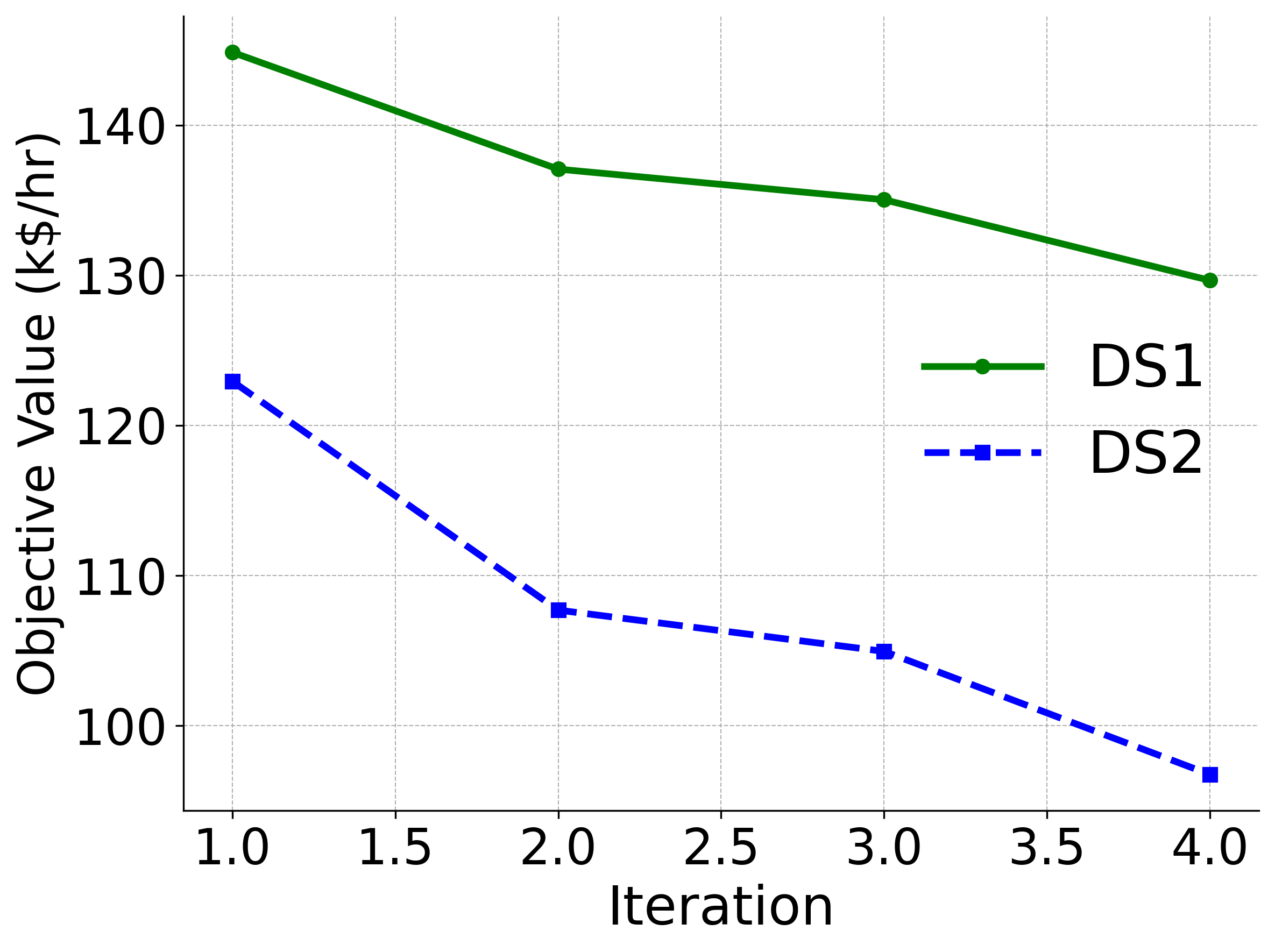}}\\
    \subfloat[IPM DCOPF]{\includegraphics[width=0.4\columnwidth]{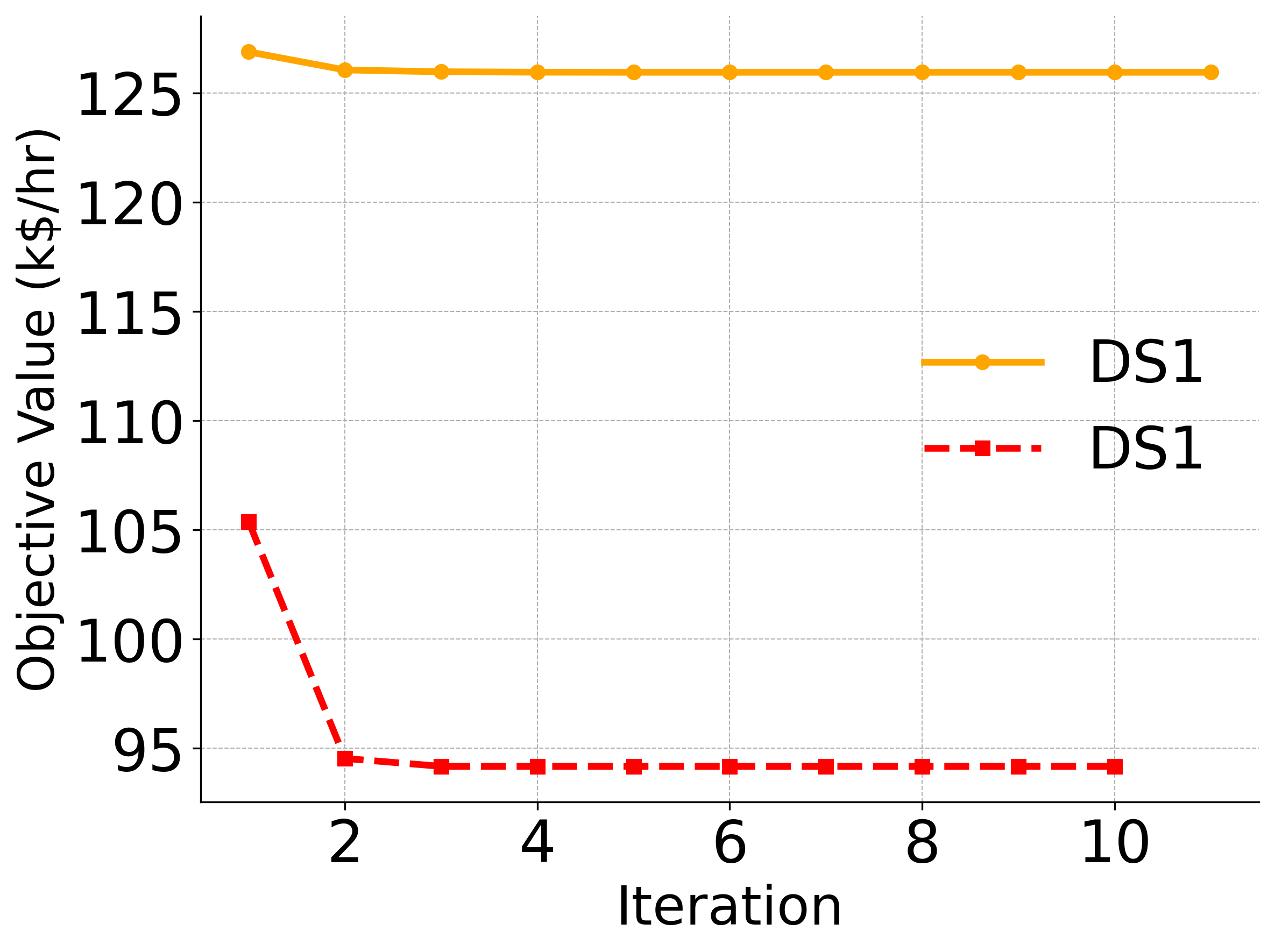}}\hfill
    \subfloat[L-IPM DCOPF]{\includegraphics[width=0.4\columnwidth]{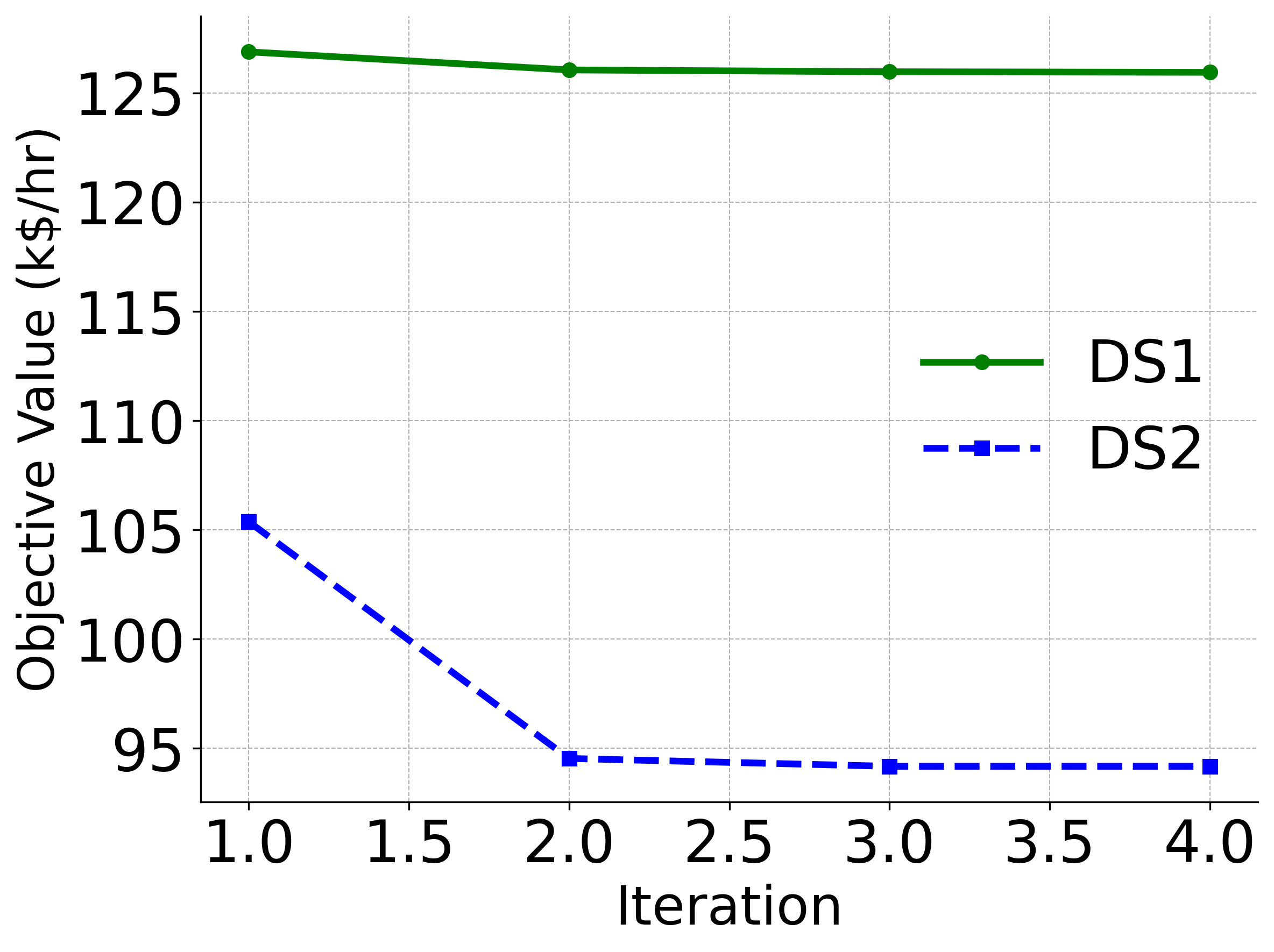}}
    \vspace{-4pt}
    \caption{OPF objective function values for the 118-bus system.}
        \vspace{-10pt}
    \label{fig:118_obj}
\end{figure}

As Figs.~\ref{fig:118_obj}(a) and (c) show, ACOPF and DCOPF exhibit stable and well-behaved convergence, even under extreme load scenarios. The system can tolerate significant load variations without any noticeable degradation in performance. The L-IPM also performed reliably, as Fig.~\ref{fig:118_obj}(b) and (d), where the results for two representative load scenarios demonstrate that the model had no difficulty in finding accurate and feasible solutions.

\subsubsection{2869-bus system}

To evaluate the scalability of our approach, we applied it to a large-scale 2869-bus system. The ACOPF formulation includes 6758 variables, along with 13,256 dual variables corresponding to IPM constraints. Due to the scale of the problem, IPM faces increased difficulty in convergence. While the method typically identifies a viable OPF solution within a few iterations, the subsequent iterations are primarily used to detect the active inequalities and enforce the physical constraints of the system.

\begin{figure}[!t]
    \centering
        \captionsetup{font={footnotesize}}
    \subfloat[IPM ACOPF]{\includegraphics[width=0.4\columnwidth]{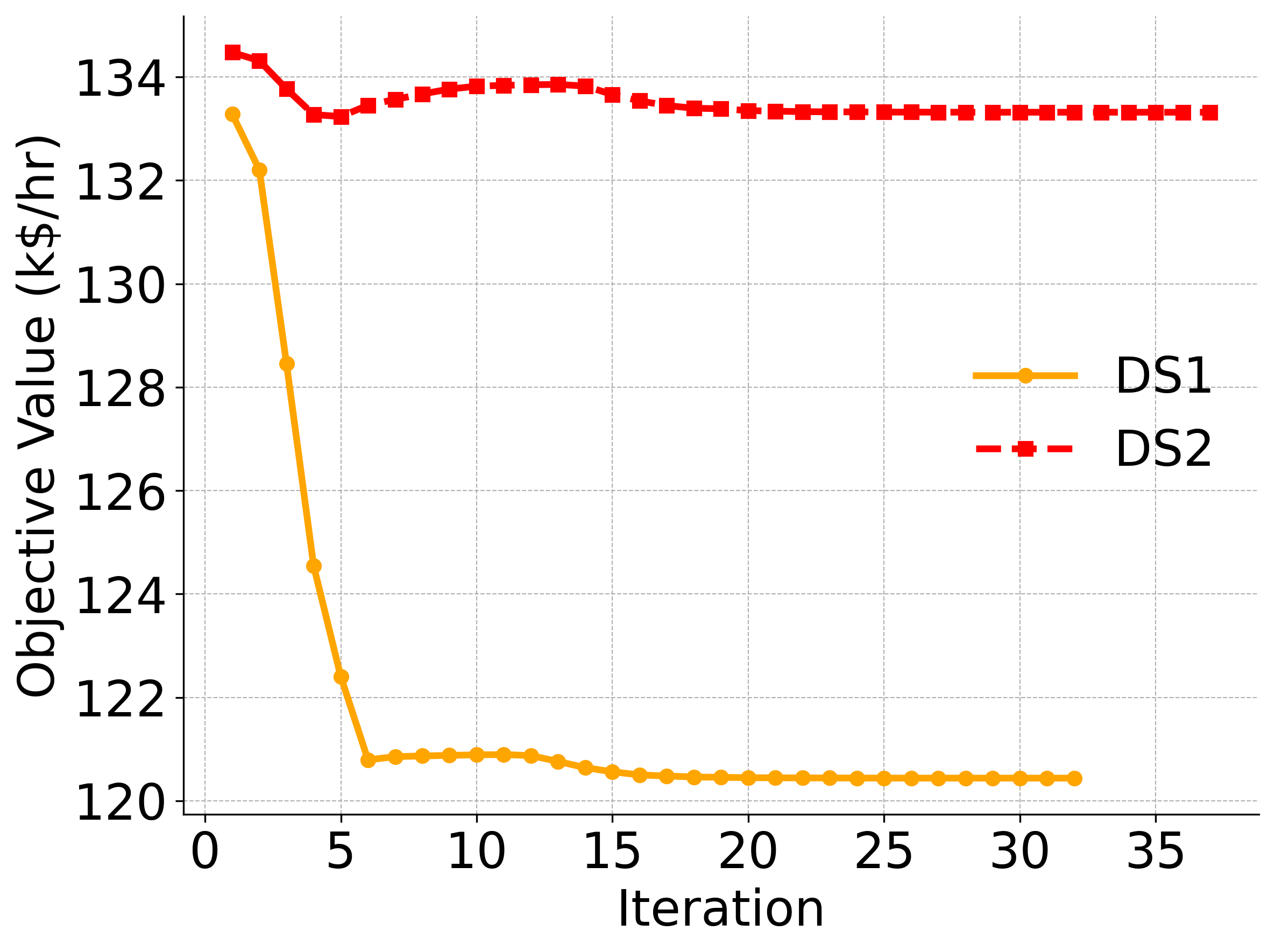}}\hfill
    \subfloat[L-IPM ACOPF]{\includegraphics[width=0.4\columnwidth]{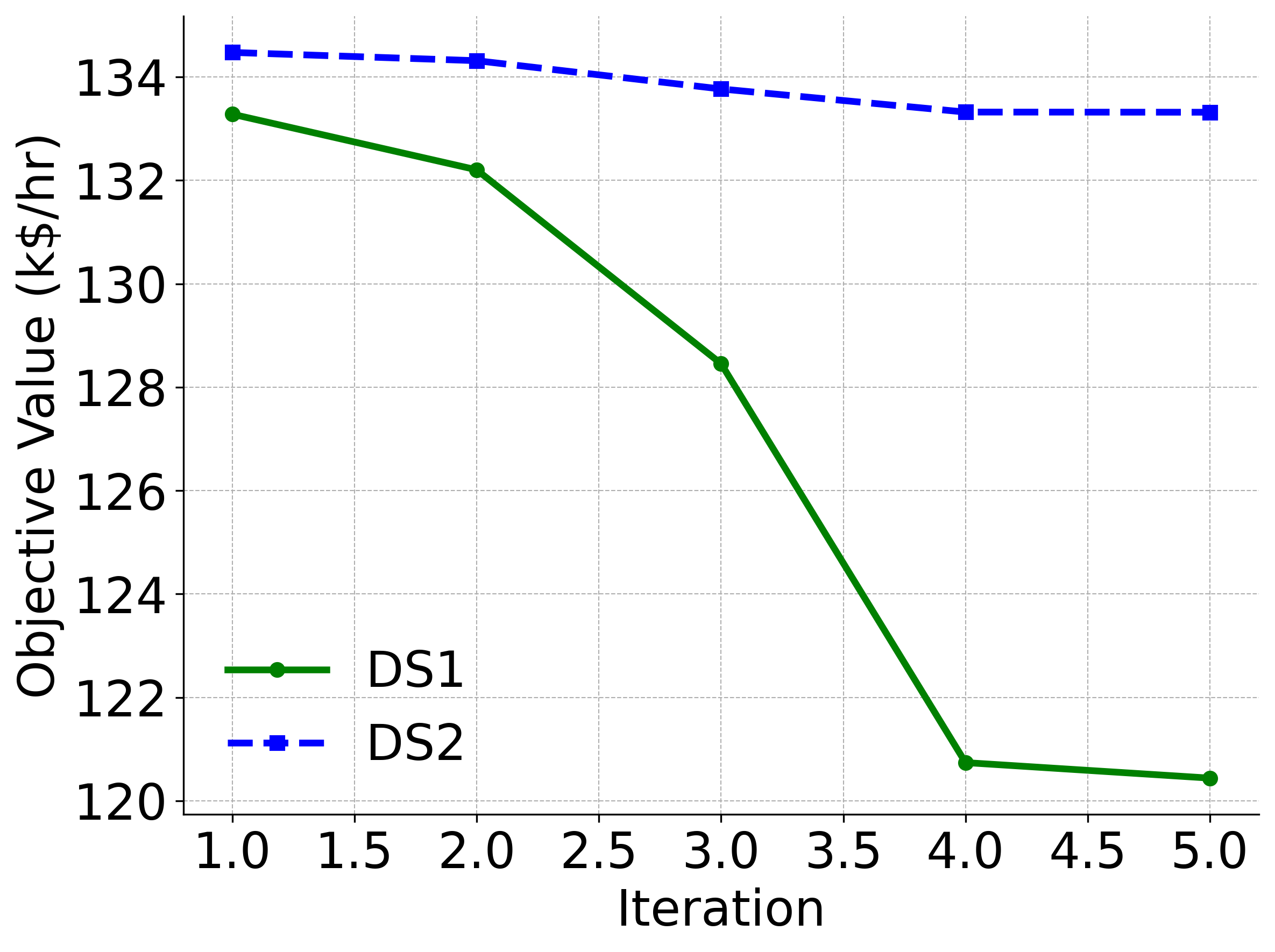}}\\
    \subfloat[IPM DCOPF]{\includegraphics[width=0.4\columnwidth]{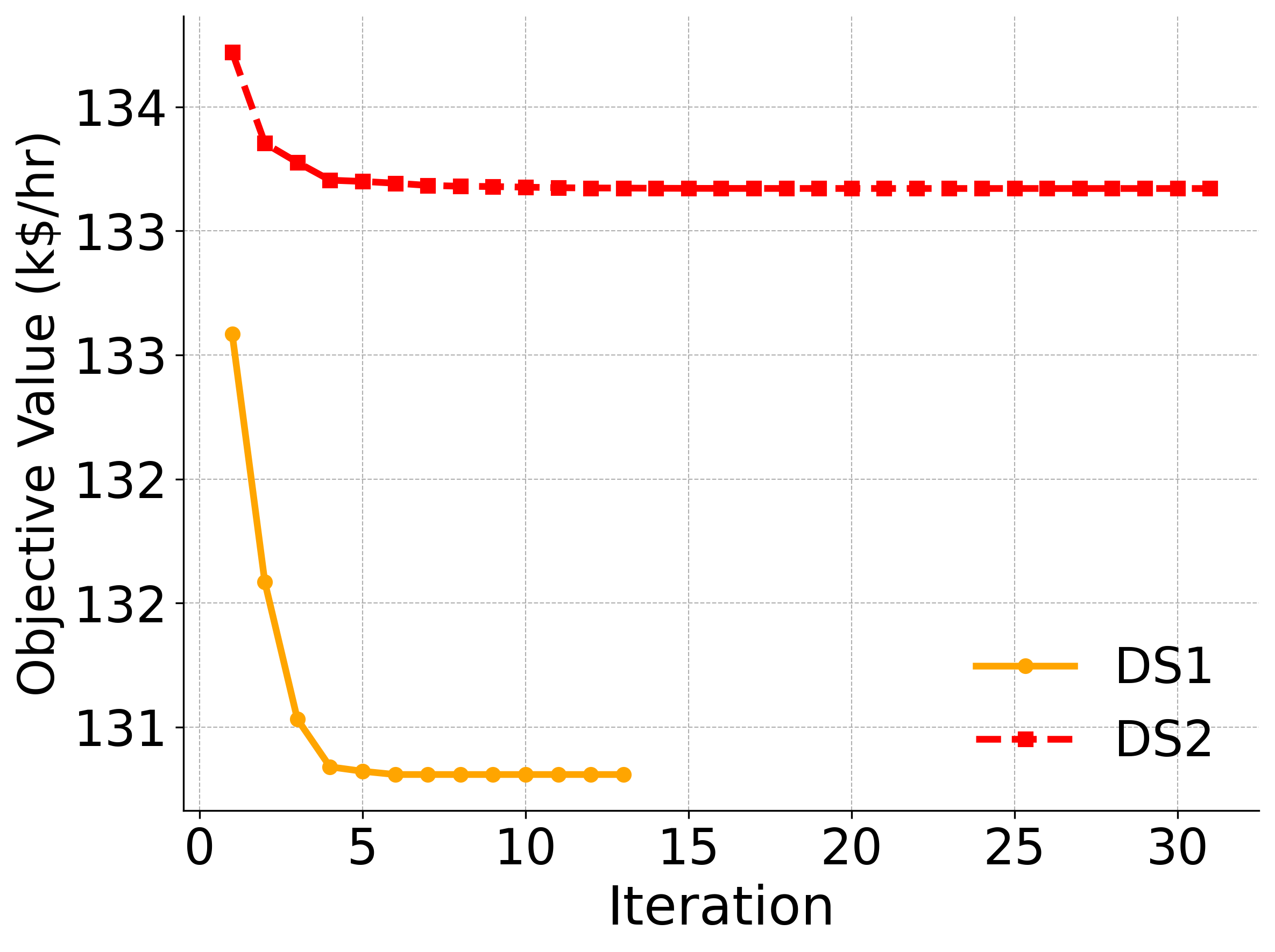}}\hfill
    \subfloat[L-IPM DCOPF]{\includegraphics[width=0.4\columnwidth]{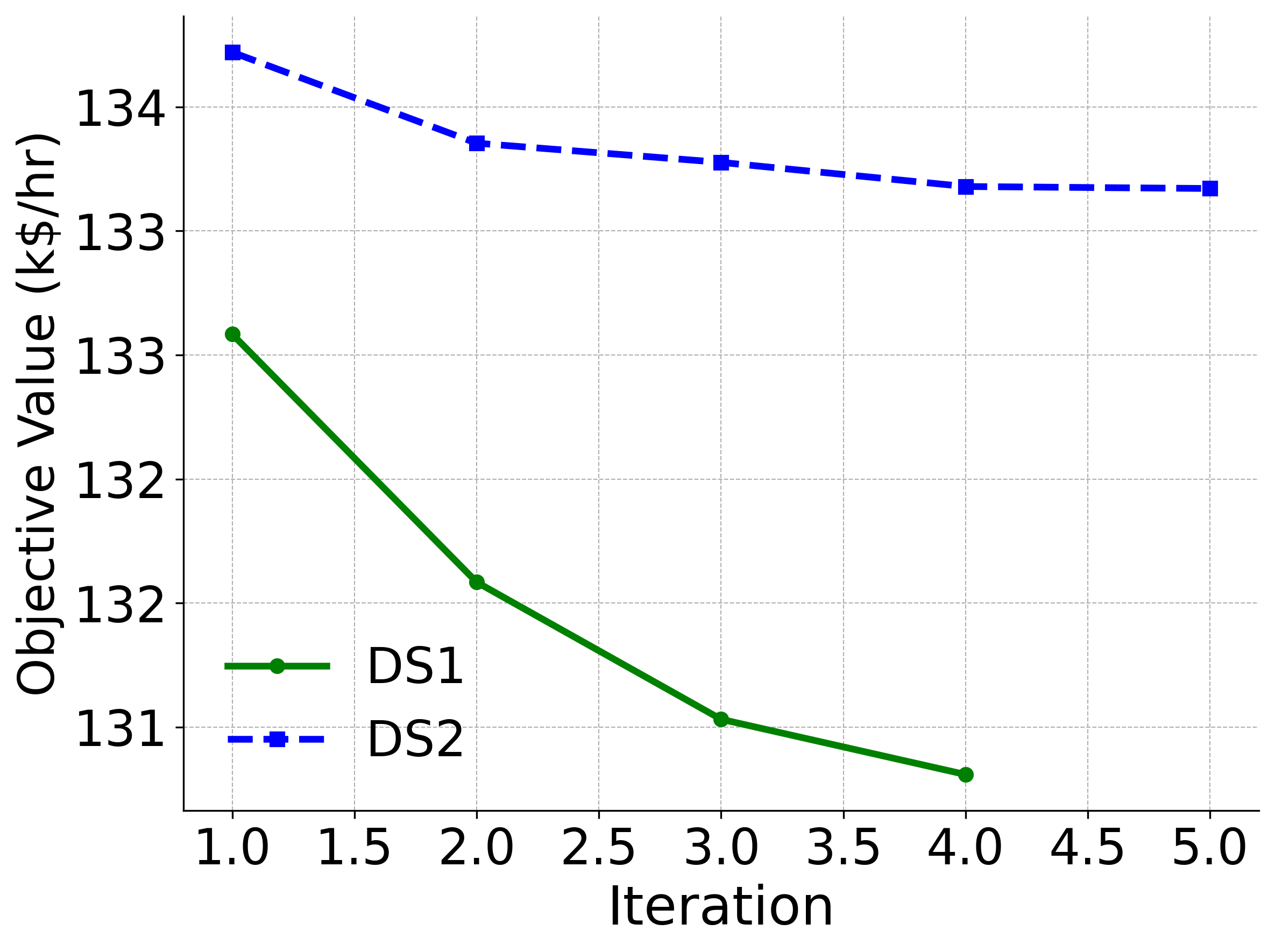}}
    \vspace{-4pt}
    \caption{OPF objective function values for the 2869-bus system.}
        \vspace{-10pt}
    \label{fig:2869_obj}
\end{figure}

\begin{figure}[!t]
    \centering
        \captionsetup{font={footnotesize}}
    \includegraphics[width=0.4\columnwidth]{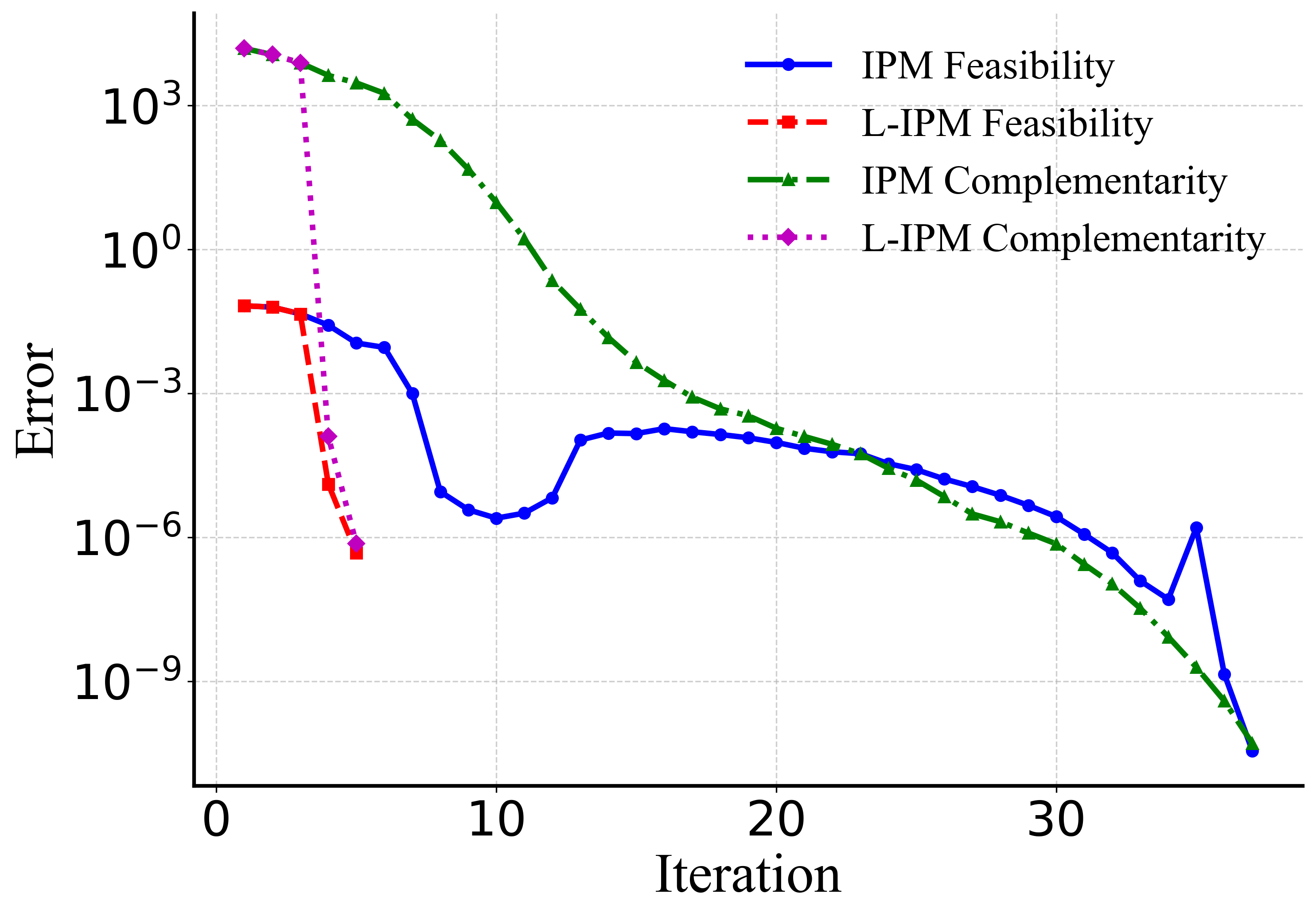}
    \vspace{-4pt}
    \caption{IPM and L-IPM Feasibility and complementarity condition comparison in 2869-bus system.}
    \vspace{-10pt}
    \label{2869FeasCom}
\end{figure}

As shown in Fig.~\ref{fig:2869_obj}, L-IPM significantly reduces the number of Newton steps required to reach convergence. Given that the dimension of the Newton linear system reaches 20,014 in this system, the computational time needed to solve each Newton step becomes substantial and directly impacts the total OPF solution time. Retaining the feasibility of the power system is critically important, as any infeasible solution can lead to severe operational and economic consequences. As shown in Fig.~\ref{2869FeasCom}, L-IPM rapidly reduces the feasibility error, demonstrating its effectiveness in guiding the solution toward a feasible and reliable operating point. 

\begin{figure}[h]
    \centering
        \captionsetup{font={footnotesize}}
    \subfloat[Objective function]{\includegraphics[width=0.43\columnwidth]{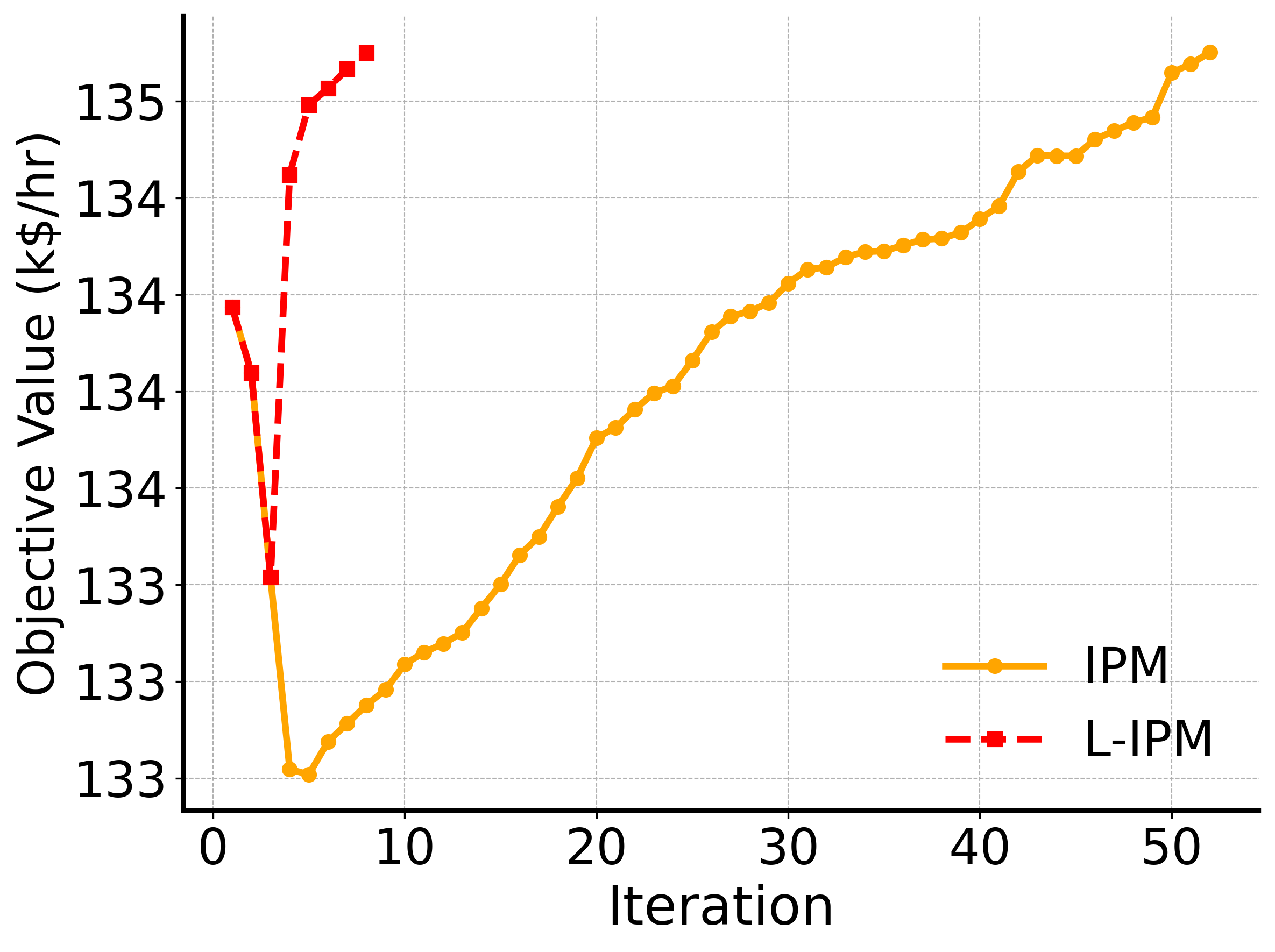}}
    \hfill
    \subfloat[Errors]{\includegraphics[width=0.4\columnwidth]{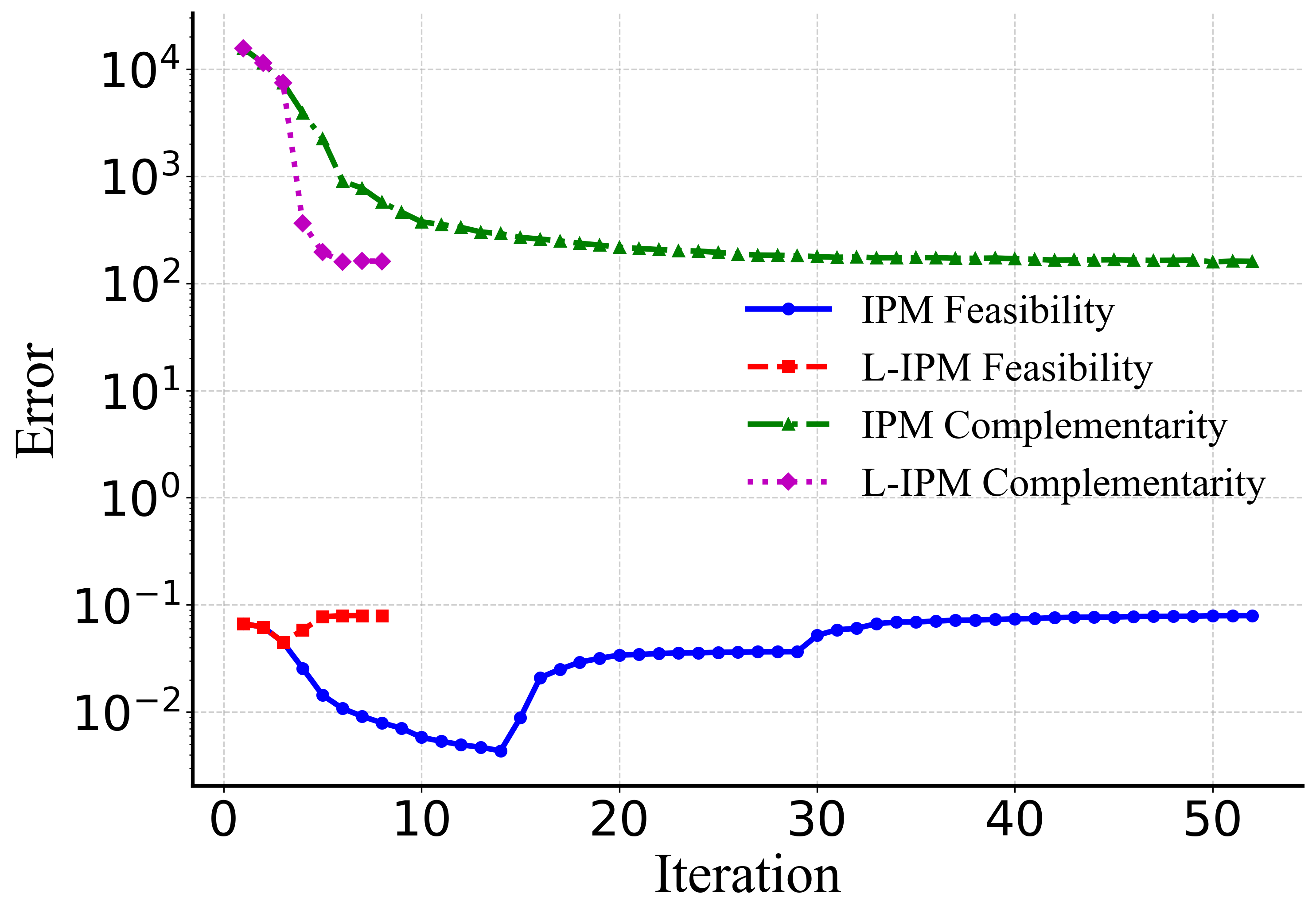}}
    \vspace{-4pt}
    \caption{IPM and L-IPM comparison in an extreme demand scenario.}
    \label{fig:infeas}
\end{figure}

\emph{Infeasible Cases Analysis}:
Given the large scale of the system, load scenarios have a significant impact on the final solution. In particular, IPM struggled to find a feasible solution under poorly conditioned load scenarios, even after many iterations. The IPM solver sometimes reached as many as 53 iterations without convergence. As illustrated in Fig.~\ref{fig:infeas}(a), we simulated a challenging load scenario where neither IPM nor L-IPM was able to converge. Even in scenarios where no feasible solution exists, L-IPM can reach this outcome more quickly than the standard IPM. When such a poorly conditioned load scenario is provided to the proposed method, the grid-informed design of GI-LSTM helps keep the predicted output within, or close to, the feasible region; however, the method correctly flags the solution as infeasible. To confirm this assessment, we examined the subsequent IPM verification step and observed that IPM fails to converge after approximately five iterations, returning a numerical failure error—consistent with the behavior of a standalone IPM run under the same scenario. The trend of the feasibility error under this scenario is presented in Fig.~\ref{fig:infeas}(b), further highlighting how the approach accelerates the detection of infeasibility while still adhering to system constraints as closely as possible.
\vspace{-4pt}

\vspace{-15pt}
\subsection{Performance of Direct Solution Prediction vs. Path-Projection}
A broad range of machine learning approaches has been explored to approximate OPF solutions by directly predicting dispatch and voltage variables through supervised learning, grid-informed neural networks, reinforcement learning, or graph-based models. In contrast, the proposed method targets the internal behavior of the OPF solver. Specifically, we leverage IPM- a widely used benchmark solver- and learn to project its central-path trajectory using informative early iterations, rather than replacing the optimizer with a black-box predictor.

A natural question is whether one can bypass the iterative process entirely by directly predicting the final OPF solution (or providing an aggressively accurate warm start). To examine the practical value of such direct prediction in the context of IPM, we conduct two stress tests: (i) initializing IPM with the exact optimal solution or 0\% neural network (NN), and (ii) initializing IPM with a perturbed solution obtained by applying a $2\%$ uniform random multiplicative error (2\% NN). 

The corresponding results are summarized in Table~\ref{table:warmstart}. It shows that, in some cases, IPM requires up to 50\% more iterations to satisfy its stopping criteria, even when initialized at the optimal primal point. This behavior is expected in a primal--dual IPM: convergence is governed not only by the primal variables, but also by the associated slack and dual variables and their complementarity/centrality conditions. Consequently, a highly accurate primal warm start alone does not necessarily place the method near the central path, and may not yield consistent computational savings.

By contrast, the proposed L-IPM framework exploits the early IPM iterations, which already encode rich information about the grid physics, active constraints, and the direction of the central path toward optimality. These early-iteration features enable effective trajectory projection and substantially improve solution quality. Finally, a concluding IPM verification step is used to assess accuracy and practicality for operational use. As shown in Table~\ref{table:warmstart}, the proposed method reduces the IPM solve to only 4 iterations across all test cases, leading to a lower runtime than the MATPOWER and NN warm-start baselines.
\vspace{-2pt}
\begin{table}[h]
\centering
\caption{Comparison of IPM iteration counts under different warm-start strategies vs. GI-LSTM.}
\vspace{-5pt}
\label{table:warmstart}
\setlength{\tabcolsep}{2.8pt} 
\renewcommand{\arraystretch}{1.05}
\resizebox{\columnwidth}{!}{%
\begin{tabular}{l cccc | cccc}
\toprule
\multirow{2}{*}{System} &
\multicolumn{4}{c|}{DCOPF} &
\multicolumn{4}{c}{ACOPF} \\
\cmidrule(lr){2-5}\cmidrule(lr){6-9}
& MATPOWER & 0\% NN & 2\% NN & GI-LSTM
& MATPOWER & 0\% NN & 2\% NN & GI-LSTM \\
\midrule
3-bus    & 10 & 13 & 13 & 4  & 11 & 16 & 16 & 4 \\
24-bus   & 25 & 29 & 30 & 4  & 13 & 17 & 17 & 4 \\
118-bus  & 11 & 14 & 15 & 4  & 14 & 18 & 19 & 4 \\
2869-bus & 12 & 18 & 18 & 4  & 36 & 45 & 46 & 4 \\
\bottomrule
\end{tabular}%
}
\vspace{1mm}
\end{table}
\vspace{-18pt}
\section{Conclusion}
This paper proposes a learning interior point method aiming to accelerate the runtime of AC and DC OPF problems. IPM finds the optimal solution by following a central path iteratively, similar to a time series. While early IPM iterations carry valuable information about the optimal solution, later ones mainly serve to ensure feasibility and are typically computationally expensive. Leveraging this, we propose a grid-informed L-IPM that reduces the number of costly iterations required to project the IPM central path and identify the optimal solution, and then verifies its feasibility using a final IPM step. 

To evaluate the proposed approach, we tested it on four systems ranging from small (3- and 24-bus) to medium and large (118- and 2869-bus) networks. These case studies demonstrate the scalability, accuracy, and computational efficiency of our method. Results show that L-IPM, when trained on carefully selected load scenarios, significantly accelerates convergence, both in runtime and in the number of IPM iterations. Under challenging load conditions, L-IPM consistently outperformed traditional IPM by requiring fewer iterations. In many cases, it reduced runtime by about 90\% compared to the classic IPM. 

Further analysis of feasibility and complementarity conditions confirmed the robustness of the proposed method, as it rapidly reduced constraint violations while maintaining solution accuracy. Although some extreme load scenarios—particularly in the 2869-bus system—introduced challenges, L-IPM still identified infeasibility more quickly than conventional IPM. A comparative analysis against classical OPF solution-prediction methods shows that initializing IPM with a high-quality primal prediction does not necessarily reduce the IPM iteration count or runtime, since IPM may still require additional iterations to properly adjust the dual variables (and barrier-related terms) and satisfy its stopping criteria. In contrast, the proposed method leverages the OPF solver’s central-path behavior to provide a more informative direction, thereby reducing the total number of IPM iterations.

Future work could focus on enhancing the model’s robustness under extreme operating conditions, incorporating component outages, and accounting for the stochastic nature of renewable energy sources and demand forecasting to improve performance under uncertainty.
\vspace{-6pt}

\bibliographystyle{IEEEtran}
\bibliography{3rd_references}

\end{document}